\documentclass[apj]{emulateapj}

\shorttitle{Tera-Leptons Shadows over  Sinister Universe}
\shortauthors{Fargion \& Khlopov}

\newcommand{\MeV}{\; \mathrm{MeV}}

\newcommand{\TeV}{\; \mathrm{TeV}}
\newcommand{\lapproxeq}{\lower .7ex\hbox{$\;\stackrel{\textstyle<}{\sim}\;$}}
\newcommand{\gapproxeq}{\lower .7ex\hbox{$\;\stackrel{\textstyle>}{\sim}\;$}}
\newcommand{\stackdown}[2]{\lower 1.4ex\hbox{$\;\stackrel{\textstyle{#1}}{\scriptstyle{#2}}\;$}}
\newcommand{\beq}{\begin{equation}}
\newcommand{\eeq}{\end{equation}}
\newcommand{\bea}{\begin{eqnarray}}
\newcommand{\eea}{\end{eqnarray}}

\def\s{{\,\rm s}}

\catcode`\@=11

\def\lsim{\mathrel{\mathpalette\@versim<}}
\def\gsim{\mathrel{\mathpalette\@versim>}}
\def\@versim#1#2{\vcenter{\offinterlineskip
    \ialign{$\m@th#1\hfil##\hfil$\crcr#2\crcr\sim\crcr } }}

\def\t1{{\tilde 1}}

\def\MeV{\,{\rm MeV}}

\def\TeV{\,{\rm TeV}}

\def\to{\rightarrow}

\def\sv{\left<\sigma v\right>}
\def\({\left(}
\def\){\right)}

\begin{document}


\title{Tera-Leptons Shadows over Sinister Universe}




%
%

\author{D. Fargion\altaffilmark{1,2}, 
  and M.Yu.~Khlopov\altaffilmark{3,4}}

\affil{\altaffilmark{1} Physics Department, Universit\'a  "La
Sapienza", Pl.A.Moro ,\altaffilmark{2} INFN, Rome, Italy}

\altaffiltext{1}{Physics Department, Universit\'a  "La Sapienza",
P.le A.Moro 5, 00185 Roma, Italy} \altaffiltext{2}{INFN Roma1,
Italy}

\affil{\altaffilmark{3} Centre for CosmoParticle Physics
"Cosmion", 125047 Moscow, Russia, \altaffilmark{4}   Moscow
Engineering Physics Institute, 115409 Moscow, Russia}

\begin{abstract}

The role of Sinister Heavy Fermions in most recent extended Glashow's
 $SU(3) \times SU(2)  \times SU(2)'  \times U(1)$  model is to offer in a unique frame relic
Helium-like products (an ingenious candidate to the dark matter
puzzle),  a solution to the See-Saw mechanism for light neutrino
masses as well as  to strong CP violation problem in QCD. The
Sinister model requires a three additional families of leptons and
quarks, but only the lightest of them Heavy $U$-quark and $E$-
\emph{electron} are stable. Apparently the final neutral
Helium-like $(UUUEE)$  state is an ideal evanescent dark-matter
candidate. However it is reached by multi-body interactions in
early Universe along a tail of more manifest secondary frozen
blocks. They should be now here polluting the surrounding matter.
Moreover, in opposition to effective $U\bar{U}$ pair annihilation,
there is no such an early or late tera-lepton pairs suppression
because: a) electromagnetic interactions are $weaker$ than nuclear
ones and b) helium ion $(^4He)^{++}$ is able to attract and
capture  (in the \emph{first three minutes}) $E^-$ fixing it into
a hybrid tera helium \emph{ion} trap. This leads to a pile up of
$(^4 HeE^-)^+$ traces, a lethal compound for any Sinister
Universe. This capture leaves no Tera-Lepton frozen in $(Ep)$
relic, otherwise an ideal catalyzer to achieve effective late
$E^+E^-$ annihilations, possibly saving the model. The $(^4
HeE^-)^+$ Coulomb screening is also avoiding the synthesis of the
desired $(UUUEE)$ hidden dark matter gas. The $(^4 HeE^-)^+e^-$
behave chemically like an anomalous hydrogen isotope. Also
tera-positronium relics $(e^-E^+)$ are over-abundant and they
behave like an anomalous hydrogen atom: these gases do not fit by
many orders of magnitude known severe bounds on hydrogen anomalous
isotope, making shadows hanging over a Sinister Universe. However
a surprising and resolver role for Tera-Pions in UHECR
astrophysics has been revealed.

\end{abstract}


\section{Introduction}
The problem of existence of new quark and lepton generations is
among the most important in the modern high energy physics. Such
new heavy leptons and quarks may be sufficiently long-living to
represent a new stable form of matter. At the present there are in
the elementary particle scenarios at least two main frame for
Heavy Stable Quarks and Leptons: (a) A fourth heavy Quark and
fourth heavy Neutral Lepton (neutrino) generation (above half the
Z-Boson mass)\cite{Shibaev}, \cite{Sakhenhance}, \cite{Fargion99},
\cite{Grossi}, \cite{Belotsky}, \cite{BKS}, \cite{Okun},
\cite{4had};  (b) A Glashow's  "Sinister" heavy quark and heavy
Charged Lepton family, whose masses role may be the dominant dark
matter . We shall briefly describe the motivation for these two
models addressing here our attention only on the Sinister Universe
suggested very recently  \cite{Glashow}.

The natural extension of the Standard model leads in the heterotic
string phenomenology to the prediction of fourth generation of
quarks and leptons \cite{Shibaev}, \cite{Sakhenhance} with a stable
massive 4th neutrino
\cite{Fargion99}, \cite{Grossi}, \cite{Belotsky}, \cite{BKS}. The
comparison between the rank of the unifying group $E_{6}$ ($r=6$)
and the rank of the Standard model ($r=4$) implies the existence
of new conserved charges and new (possibly strict) gauge
symmetries. New strict gauge U(1) symmetry (similar to U(1)
symmetry of electrodynamics) is possible, if it is ascribed to the
fermions of 4th generation. This hypothesis explains the
difference between the three known types of neutrinos and neutrino
of 4th generation. The latter possesses new gauge charge and,
being Dirac particle, can not have small Majorana mass due to see
saw mechanism. If the 4th neutrino is the lightest particle of the
4th quark-lepton family, strict conservation of the new charge
makes massive 4th neutrino to be absolutely stable. Following this
hypothesis \cite{Shibaev} quarks and leptons of 4th generation are
the source of new long range interaction, similar to the electromagnetic
interaction of ordinary charged particles.

Recent analysis \cite{Okun} of precision data on the Standard
model parameters has taken into account possible virtual
contributions of 4th generation particles. It was shown that 4th
quark-lepton generation is not excluded if 4th neutrino, being
Dirac and (quasi-)stable, has a mass about 50 GeV (47-50 GeV is
$1\sigma$ interval, 46.3-75 GeV is $2\sigma$ interval) \cite{Okun}
and other 4th generation fermions satisfy their direct
experimental constraints (above 80-220 GeV).

Heavy electron of 4th generation, $E^-$, is in this case unstable
and decays due to charged current weak interaction to $E
\rightarrow N+W$, where, depending on the mass ratio of $E^-$ and
$N$, $W$ boson is either real or virtual.

However a very novel Glashow's  "Sinister" $SU(3)_c \times SU(2)
\times SU(2)' \times U(1)$ gauge model \cite{Glashow} offers
another possible realization for the heavy Quark - Lepton
generations. Three Heavy generations of tera-fermions are related
with the light fermions by $CP'$ transformation linking light
fermions to charge conjugates of their heavy partners and vice
versa. $CP'$ symmetry breaking makes tera-fermions much heavier
than their light partners. Tera-fermion mass pattern is the same
as for light generations, but all the masses are multiplied by the
same factor $S =10^6 S_6 \sim 10^6$. Strict conservation of $F =
(B-L) - (B' - L')$ prevents mixing of charged tera-fermions with
light quarks and leptons. Tera-fermions are sterile relative to
SU(2) electroweak interaction, and do not contribute into standard
model
parameters. 
 In such realization the new heavy neutrinos ($N_i$)
  acquire large masses and their
mixing with light neutrinos $\nu$ provides a "see-saw" mechanism of
light neutrino Dirac mass generation. Here in a Sinister model the
heavy neutrino is unstable. On the contrary in this scheme $E^-$ is
the lightest heavy fermion and it is absolutely stable.
If the lightest quark $Q$ of Heavy generation does not mix with
quarks of 3 light generation, it can decay only to Heavy
generation leptons owing to GUT-type interactions, what makes it
sufficiently long living. If its lifetime exceeds the age of the
Universe\footnote{If this lifetime is much
less than the age of the Universe, there should be no primordial
Heavy generation quarks, but they can be produced in cosmic ray
interactions and be present in cosmic ray fluxes
 or in their most
recent relics on Earth. The assumed masses for these
tera-particles make their search a challenge for the present and
future accelerators.}, primordial $Q$-quark hadrons
as well as Heavy Leptons
should be present in the modern matter\footnote{The mechanisms of production of metastable $Q$ (and $\bar Q$)
hadrons in the early Universe, cosmic rays and accelerators were
analyzed in \cite{4had} and the possible signatures of their wide
variety and existence were revealed. Such hadrons were assumed to
be bound states $(Qqq)$ of heavy $Q$ and light quarks $q$  formed
after QCD confinement.}.

Indeed,
in the novel Glashow's "Sinister" scenario \cite{Glashow}
 very heavy quarks $Q$ (or antiquarks $\bar Q$) can form bound states with other heavy quarks (or antiquarks) due to their Coulomb-like QCD attraction, and the binding energy of these states may substantially exceed the binding energy of QCD confinement.
Then $(QQq)$ and $(QQQ)$ baryons can exist.

In the model \cite{Glashow} the properties of heavy generation
fermions were fixed by their discrete $CP'$ symmetry with light
fermions. According to this model heavy quark $U$ and heavy
electron $E$ are stable and
may form a neutral most probable and stable (while being evanescent) $(UUUEE)$ "atom"
with $(UUU)$ hadron as nucleus and two $E^-$s as "electrons". The
tera gas of such "atoms" is an ideal candidate for a very new and fascinating dark matter;
because of their peculiar WIMP-like interaction with matter they may also rule the  stages of gravitational clustering in
early matter dominated epochs, creating first gravity seeds for galaxy formation.

 However, while the assumed terabaryon asymmetry for
$U$ washes out by annihilation primordial $\bar U$, the tera-lepton asymmetry of $E^-$ can not effectively suppress the abundance of tera-positrons $E^+$ in the earliest as well as in the late
Universe stages. This feature differs from successful and most celebrated annihilation of primordial antiprotons and
positrons that takes place in our Standard baryon asymmetrical Universe. The abundance  of $\bar U$ and $E^+$
in earliest epochs exceeds the abundance of excessive
$U$ and $E^-$ and it is suppressed (successfully) for $\bar U$ only
after QCD phase transition,
while, as we shall stress, there is no such effective annihilation
mechanism for $E^+$. Moreover ordinary $^4$He formed in Standard
Big Bang Nucleo-synthesis binds at $T \sim 15 keV$ virtually all
the free $E^-$ into positively charged $(^4HeE^-)^+$ "ion", which
puts Coulomb barrier for any successive $E^-E^+$ annihilation or
any effective $EU$ binding.
 The huge frozen abundance of tera-leptons in hybrid tera-positronium $(eE^+)$
 and hybrid hydrogen-like tera-helium atom $(^4He Ee)$ and
 in other complex anomalous isotopes can not be removed\footnote{Anyway the eventual energy release of this late $E E^+$
annihilation, if it was possible, can inject energy and cause
distortions in CMB spectrum, as well as influence the relic
products in nuclei formed in SBBN  changing the light element
abundance. These bounds are less restrictive but as we stressed
are overcome by Helium trap capture.}.



Indeed in analogy to D, $^3$He and Li relics that are the
intermediate catalyzers of $^4$He formation
the tera-lepton and tera-hadron relics from
intermediate stages
towards a final $(UUUEE)$
formation must survive with high abundance of {\it visible} relics in the present Universe.
We  enlist, reveal and classify such tracers, their birth
place and history up to now; we shall remind their lethal role for
the present and wider versions of Sinister Universe  \cite{Glashow}. We find that $(eE^+)$
and $(UUUEe)$,
which we label hybrid tera-hadron atom, is here to remain among us and their
abundance is
enormously high for known severe bounds on anomalous hydrogen. Moreover evanescent relic neutral hybrid tera-hydrogen atom $(Ep)$ can not be formed
because the primordial component of free tera-electrons $E^-$ are
mostly trapped in the first minutes into hybrid hydrogen-like
tera-helium ion $(^4HeE^-)^+$, a surprising cage, screening by
Coulomb barrier
any eventual later $E^+E^-$ annihilation or $EU$ binding. This is the grave nature of tera-lepton shadows over a Sinister Universe.
Therefore remaining abundance of $(eE^+)$ and $(^4HeE^-e)$ exceeds by  {\it 27 orders}
of magnitude the terrestrial upper limit for anomalous hydrogen. There are also additional tera-hadronic anomalous relics, whose trace is constrained by the present data by  {\it 25.5 orders} for $(UUUEe)$ and {\it at least by 20 orders} for $(Uude)$ respect to anomalous hydrogen, as well as by {\it 14.5 orders} for $(UUUee)$,
by {\it 10 orders} for $(UUuee)$ and $(Uuuee)$ (if $(Uuu)$ is the lightest tera-hadron) - respect to anomalous helium.
While tera helium $(UUUEE)$ would co-exist with observational data, being a wonderful candidate for dark matter, its tera-lepton partners will poison and forbid this opportunity.

The contradiction might
be removed, if tera-fermions are unstable and drastically decay before the
present time. But such solution excludes any cosmological sinister
matter dominated Universe, while, of course, it leaves still room and challenge for
search for metastable $E$-leptons and $U$-hadrons in laboratories or in High Energy Cosmic ray traces.

The paper is structured as follows. We discuss possible types of
stable tera particles in Section 2 stipulate their
early chronological and thermal history in Section
3; here we show that while it is possible to
suppress primordial anti-tera-quarks by pair annihilation, the
similar annihilation is inhibited  for tera-leptons. Consequent
formation of tera-helium $(UUUEE)$ is inevitably accompanied by
dominant fraction of charged tera-leptons; while this lepton
component is sub-dominant in energy density still it is
over-abundant for known bounds even at smallest possible $S$
values. In Section 4 we follow at $T \sim 15 keV$ free
$E^-$ binding with $^4$He, trapped into a mortal cage of
positively charged $(E ^4He)^+$ ion. This capture precludes
formation of neutral $(E^-p)$ and leaves no room to a hopeful
proposal \cite{Glashow} of $(Ep)$ catalytic elimination of all the
products of "incomplete tera-matter combustion". In the result all
the positively charged tera-matter fragments recombine with
ordinary electrons into over-abundant fraction of anomalous
isotopes both of tera-leptons ($(eE^+)$ and $(^4HeE^-e)$) as well
as of  tera-hadron ($(UUU)^{++}$ (pure tera-helium ion),
$(UUu)^{++}$ (first hybrid tera-helium ion),
 $(Uuu)^{++}$ (second hybrid tera-helium ion), $(UUUE^-)^{+}$ (tera-helium I ion) ,
 $(UUuE^-)^{+}$ (first hybrid tera-helium I ion), $(UuuE^-)^{+}$) nature, which can not be effectively suppressed
to present upper limits of anomalous isotopes by any realistic
mechanism (Section 5). A summary  is found in last
Section 6 and Section 7.


\section{\label{lightest} The lightest stable Heavy hadrons in Bounded Atoms}

In the framework of the "sinister" $SU(3) \times SU(2)  \times SU(2)'  \times U(1)$ gauge model \cite{Glashow} quarks of heavy generations follow the same mass hierarchy as
their light partners but differ by a factor $S = 10^6 \cdot S_6$. It makes $U$ quark with
mass $S m_u = S \cdot 3.5 MeV$ the lightest Heavy quark.
In principle, the composition of the lightest  Heavy baryons can be: $(Uuu)$ (charge $+2$)\footnote{One can call
this particle $\Delta^{++}_{U}$ or $\Sigma^{++}_{U}$.}, $(Uud)$
(charge $+1$), $(Udd)$ (charge $0$), $(UUu)$
(charge $+2$) and the
corresponding lightest mesons $(\bar U u)$ (charge $0$) or $(\bar U
d)$  (charge $-1$). Charmonium-like $(\bar U U)$ state is unstable
relative to 2- or 3-gluon decays.

Because of the heavy "lightest" Quark masses considered here form
the most deep binding nuclear potential,  the final
$(UUU)^{++}$ (says also $\Delta^{++}_{UUU}$) state is the most
stable to be formed in early Universe. This state occurs via
interfaced states $Uqq$ and $UUq$, where $q$ are light quarks.

The QCD phase transition for "light" quarks took place at few hundred MeV but
the binding gluon energy is related to the Quark masses scaled by
whose coupling in imagined $10^6 \cdot S_6$ (as $\frac{m_U}{m_u}$ or
$\frac{m_E}{m_e}$ ratio masses ), leading to a binding hadron
energy at $\alpha_{QCD}^2
M_{Quark}/4 \simeq 1.5 \cdot 10^{10} eV S_6 $, larger by two orders
of magnitude respect common hadrons.

The minimal value of $S$ factor $S_6 = 0.2$ follows \cite{Glashow} from unsuccessful search for heavy leptons with mass below $100 GeV$.
The heavy electron masses at energy $m_E \simeq 10^6 S_6 m_e \simeq
500 GeV S_6$ are leading to a
$(UUU)EE$
anomalous Atom, whose binding energy may reach $E_{Binding}\simeq
Z^2 \alpha^2 M_E/2 \sim 5 \cdot 10^7 eV S_6$; because of it these atoms are quite
stable and bounded, while interacting only by multi-pole
electromagnetic states.

The prediction  \cite{Glashow} that
$m_D>m_U$ excludes stable baryon with the negative charge,
while the lightest $(\bar U d)$ meson seems to be excluded by the
quark model arguments. These arguments also exclude neutral
$(Udd)$ as the lightest single-$U$ baryon, as well as $(UUd)$
as the lightest double-$U$ baryon.

It leaves theoretically favorable $(Uud)$ and theoretically less
favorable $(Uuu)$ as only candidates for lightest single $U$-quark baryon\footnote{The argument favoring $(Uuu)$ as the lightest $U$ baryon was simply
based in  \cite{4had} on the mass ratio of current $u$ and $d$ quarks. On the
other hand there is an argument  \cite{4had} in favor of the fact that the
$Uud$ baryon must be lighter than the $Uuu$ one.
 Indeed, in all models the scalar-isoscalar $ud$-diquark is lighter
than the vector-isovector $uu$-diquark. One example is the model
with the effective t'Hooft instanton induced four quark
interaction, which provides a rather strong attraction in the
scalar $ud$-channel and which is absent in the vector $uu$
channel. Thus it is very likely that the $Uuu$-baryon will be unstable
under the decay $(Uuu)\to (Uud)\ +\  \pi^+$. This expectation \cite{4had} is
confirmed  by the properties of the charmed baryons (where the
charm quark is much heavier than two other quarks). Indeed, the
branching of the $\Sigma_c\to \Lambda_c +\pi$
decay is about 100\% \cite{PDG}.
In a more general form we can say \cite{4had} that the
interaction in isoscalar channel (isoscalar potential) must be
stronger than that in the isovector case. Otherwise we will obtain
a negative cross section for one or another reaction since the
isovector interaction changes the sign under the replacement of
$d$-quark by the $u$-quark.},
$(UUu)$ and $(UUU)$ as the lightest multi-$U$-quark baryons and with the only possibility of neutral $(\bar U u)$ (and it antiparticle $(U \bar u)$) as the lightest meson.
In the present paper we
 choose for our estimates the value $M_U=3.5 S_6$ TeV.

\section{\label{primordial} Primordial Tera-particles from Big Bang Universe}
The model  \cite{Glashow} assumes that in the early Universe
charge asymmetry of tera-fermions was generated so that $UUU$ and $EE$
excess saturated the modern dark matter density. For light baryon excess
$\eta_b= n_{b\,mod}/n_{\gamma \,mod} = 6 \cdot 10^{-10}$ it gives tera-baryon excess \cite{Glashow}
\beq
\eta_{B'} = 3 \cdot 10^{-13} (\frac{3.5{\TeV}}{m}),
\label{excess}
\eeq
where $m$ is the mass of $U$-quark. For future use it is convenient to relate baryon $\Omega_b=0.044$
and tera-baryon number densities $\Omega_{CDM}=0.224$ \cite{Glashow} with the entropy density $s$ and to introduce
$r_b = n_b/s$ and $r_{B'}=n_{B'}/s$. Taking into account that $s_{mod}=7.04\cdot n_{\gamma\,mod},$ one obtains $r_b \sim 8 \cdot 10^{-11}$ and
\beq
r_{B'} = 4 \cdot 10^{-14} (\frac{3.5{\TeV}}{m}).
\label{sexcess}
\eeq
It is assumed in \cite{Glashow} that $B'=1/3$ for $U$ quark,
so the $B'$ excess Eq.(\ref{sexcess}) corresponds to $U$ quark excess $r_U$
given by
$$\kappa_{U} =r_{U} -r_{\bar U} = 1.2 \cdot 10^{-13} (\frac{3.5{\TeV}}{m}) =$$
\beq
 = 1.2 \cdot 10^{-13}/S_6,
\label{Uexcess}
\eeq
where $S_6 = S/10^6$.
To have equal amounts of $UUU$s and $EE$s one needs two tera-leptons
per one tera-baryon. It means that $E$ excess should be equal to
$$r_{L'} = r_{E}- r_{E^+}  = 2 \cdot r_{B'} = 8 \cdot 10^{-14} (\frac{3.5{\TeV}}{m}) = $$
\beq
= 8 \cdot 10^{-14}/S_6.
\label{Eexcess}
\eeq
The excess Eq.(\ref{Eexcess}) corresponds to the nonzero $B'-L'$.
From the conservation of $F = (B-L) - (B' - L')$, assumed in \cite{Glashow},
it implies nonzero $B-L$ in light baryon excess generation.
The latter preserves light baryon asymmetry from washing out
due to electroweak baryon number nonconservation,
which makes $B+L=0$.

The goal of the successive discussion is to reveal the possible
tracers from various steps of cosmological evolution of primordial $U$ and $E$.
The main idea of this treatment is that if some particles of type $1$ and $2$
form in the early Universe a product $3$, the process of this transformation freezes out in the period $t_r$, when the rate of expansion exceeds the rate of reaction $1+2 \rightarrow 3$.
For comparable relative abundance (in terms of entropy, $s$) of $1$ and $2$ ($r_1=n_1/s \sim r_2=n_2/s \gg k = (n_1-n_2)/s$),
their frozen out abundance is $r_1\approx r_2 \sim 1/(s \sigma v t_r) = 1/J$, while if some particles are in excess and in the period of freezing out $r_1=k \gg r_2=r$ the amount of excessive particles practically does not change ($r_1 = k$), while the amount of $2$ becomes exponentially small $r_2 \sim r \exp{(-k s \sigma v t_r)} = r \exp{(-k J)}$. We show that since abundance of $U$ and $E$ is comparable and effects of tera-baryon asymmetry are in many cases not too strong, significant amount of "incomplete combustion" products should exist in the Universe for a sufficiently long period and even survive to the present time.
\subsection{\label{Chronology} Chronological cornerstones of Sinister Universe}
After generation of tera-baryon and tera-lepton asymmetry in chronological order thermal history of tera-matter looks as follows for $m_U = 3.5 S_6$TeV and $m_E = 500 S_6$ GeV

1) $10^{-13}\s/S_6^2 \le t \le  10^{-10}\s/S_6^2$ at $m_U\ge T \ge T_U=m_U/30 \approx 100 GeV S_6$
Tera-quark pairs annihilation and freezing out, leaving the earliest non-negligible abundance of $U \bar U$ pairs (Subsection \ref{Ufreezing} and Appendix 1).

2) $4 \cdot 10^{-12}\s/S_6^2 \le t \le 2.5 \cdot10^{-10}\s/S_6^2$  at $m_E\ge T \ge T_E=m_E/25 \approx 20 GeV S_6$
Tera-lepton pair $E \bar E^+$ annihilation and freezing out (Subsection  \ref{Efreezing} and Appendix 1).

3) $ t \sim 4.5 \cdot10^{-10}\s/S_6^2$  at $T \sim I_U= \alpha^2_c m_U/4 \approx 15 GeV S_6.$
At this temperature, corresponding to $U$-quark chromo-Coulomb binding energy $I_U \approx 15 GeV S_6$ binding of $\bar U$ in "tera-charmonium" $(U \bar U)$ and their immediate annihilation takes place. This process is very effective to suppress most of $U \bar U$ pairs (Appendix 3).

4) $4.5 \cdot 10^{-9}\s/S_6^2 \le t \le 4 \cdot10^{-6}\s/S_6^2$  at $I_U\ge T \ge I_U/30 \approx 0.5 GeV S_6$
Binding of $U$-quarks in $UU$- diquarks and $(UUU)$-hadrons  (Appendix 4).

5) $ t \sim 4.5 \cdot10^{-5}\s$  at $T \sim T_{QCD}=150 MeV.$
QCD phase transition. $UU$ recombination and $\bar U U$ annihilation in hadrons (Appendix 5)\footnote{After QCD confinement $U$-quarks and $UU$- diquarks form hybrid tera-hadrons $(Uuu)$ (or $(Uud)$) and $(UUu)$, while $\bar U$ form $(\bar U u)$.
Hadronic recombination provides binding of $U$-quarks and diquarks in $(UUU)$
in reactions like $(UUu)+(Uud) \rightarrow (UUU) + hadrons$. Anti-tera quark hadrons are additionally suppressed in processes  $(Uud) + (\bar U u) \rightarrow (U \bar U)+ hadrons \rightarrow hadrons$.}.

6) $4 \cdot 10^{-4}\s/S_6^2 \le t \le 4.5 \cdot10^{-1}\s/S_6^2$  at $I_{UE}\ge T \ge I_{UE}/30 \approx 1.5 MeV S_6.$ In this period tera-electron $E^-$ recombination with positively charged  $U$-hadrons and tera-helium "atom" $(UUUEE)$ formation with potential energy
$I_{UE}= Z^2 \alpha^2 m_E/2 \approx 50 MeV S_6$ ($Z=2$) takes place (subsection \ref{recE})\footnote{Together with $(UUUEE)$ also $(UUUE)^+$, $(UUuE)^+$, $(UuuEE)$ and $(UuuE)^+$ (or $(UudE)$ are formed, while
free $E^-$, $(UUU)^{++}$, $(UUu)^{++}$, $(Uuu)^{++}$ (or $(Uud)^+$) are left.}.

7) $ t \sim 2.8 \cdot10^{-2}\s/S^2_6$  at $T \sim I_{E^+} =6.3 MeV S_6.$
The temperature corresponds to binding energy $I_{E^+}=  \alpha^2 m_E/4 \approx 6.3 MeV S_6$ of twin tera-positronium $(E^-E^+)$ tera-atom, in which $E^+$ annihilate.
This annihilation contrary to earlier $\bar U U$ binding (see p.3) is not at all effective to reduce the $E^-E^+$ pairs abundance (subsection \ref{anE}). This is one of the main reasons, why Sinister Universe is not compatible with observations.

8) $100\s \le t \le 4.5 \cdot10^{3}\s$  at $100 keV\ge T \ge I_{EHe}/27 \approx 15 keV,$
where $I_{EHe}= Z^2 \alpha^2 m_{He}/2 = 400 keV$ is the potential energy of both $(^4HeE^-E^-)$ atom and of $(^4HeE^-)^+$ ion.
Helium $^4$He is formed in Standard Big Bang Nucleosynthesis and virtually all free $E^-$ are trapped by $^4$He in $(^4HeE^-)^+$ ion (section 4)\footnote{Note that in the period $100 keV \le T \le 400keV$ helium $^4$He is not formed, therefore it is only after {\it the first three minutes}, when lethal $(^4HeE^-)^+$ trapping of $E^-$ can take place. Coulomb barrier inhibits successive reaction of $E^-$ with positively charged tera-particles and $E^+$, $(^4HeE^-)^+$, $(UUUE)$, $(UUuE)$, $(UuuE)$, $(UUU)$, $(UUu)$, $(Uuu)$ (or $(Uud)$) can no more decrease their abundance.}.

9) $1.6 \cdot 10^{3}\s \le t \le 10^{6}\s$  at $I_{Ep}\ge T \ge I_{E}/25 \approx 1 keV.$
Aborted $(Ep)$ capture because of earlier $(^4HeE^-)^+$ trapping of free $E^-$. Here $I_{Ep}=  \alpha^2 m_{p}/2 = 25 keV$ is the potential energy of hypothetical $(Ep)$ atom and $T \approx I_{Ep}/25=1 keV$ would correspond to the end of $(Ep)$ binding (section 4)\footnote{A tedious reader would argue that within $1.6 \cdot 10^{3}\s \le t \le 4.5 \cdot 10^{3}\s$ both $(^4HeE^-)^+$ and $(Ep)$ capture could take place, but the most of $E^-$ are much faster trapped in $(^4HeE^-)^+$, than by any late $(Ep)$ binding.}.

10) $t \sim 2.5 \cdot 10^{11}\s$  at $T \sim I_{He}/30 \approx 2 eV.$
Here $I_{He}= Z^2 \alpha^2 m_{e}/2 = 54.4 eV$ is the potential energy of ordinary He atom.
Formation of anomalous helium atoms (Appendix 6 and section 5)\footnote{Relics with charge $Z=+2$ recombine with $e^-$ and form anomalous helium atoms $(UUUee)$, $(UUuee)$, $(Uuuee)$ (if $(Uuu)$ is lightest).}.

11) $z \sim 1500.$ Last scattering and common hydrogen recombination are accompanied by
anomalous positronium and hydrogen atoms formation(Appendix 6 and section 5)\footnote{Relics with charge $Z=+1$ recombine with $e^-$ and form anomalous hydrogen atoms $(^4HeE^-)^+e^-$, $(E^+ e^-)$, $(UUUEe)$, $(UUuEe)$, $(UuuEe)$ (or $(Uude)$).}.

All these anomalous species should be present in matter around us and we turn now to the stages of their formation.
\subsection{\label{Ufreezing} Freezing out of $U$-quarks}

In the early Universe at temperatures highly above their masses
tera-fermions  were in thermodynamical equilibrium
with relativistic plasma. It means that at $T>m$ the excessive $E$ and $U$ were
accompanied by $E E^+$ and $U \bar U$ pairs.

When in the course of expansion
the temperature $T$ falls down
\footnote{This picture assumes that reheating temperature $T_r$
after inflation exceeds $m$. A wide variety of inflationary models
involve long pre-heating stage, after which reheating temperature
does not exceed $T_r < 4\cdot 10^6 GeV$. This upper limit appears,
in particular, as necessary condition to suppress over-abundance
of primordial gravitino (see e.g. \cite{Linde},  for review and
Refs. \cite{book}). Therefore the successive description of
freezing out of tera-quarks may not be strictly applicable for
very large $S_6 > 10^3$, when nonequilibrium mechanisms of
tera-particle creation can become important. However, even the out
-of-equilibrium mechanisms of tera-particle creation in early
Universe can hardly avoid appearance of tera-quark and tera-lepton
pairs.}
below the mass of $U$-quark, $m$, the concentration of quarks and
antiquarks is given by equilibrium.
At the freezing out temperature $T_f$ the rate of expansion
exceeds the rate of annihilation to gluons $U \bar U \rightarrow
gg$ or to pairs of light $q$ quarks  and $\bar q$ antiquarks $U \bar U \rightarrow
\bar q q$. Then quarks $U$ and antiquarks $\bar U$ are frozen out.

The frozen out
concentration (in units of entropy density) of $U$ quarks, $r_{U}$, and antiquarks, $r_{\bar U}$,
is given  (see Appendix 1) by
\begin{eqnarray}
r_U=8.6 \cdot 10^{-13} f_{U} (S_6)\nonumber\\
r_{\bar U}= 7.4 \cdot 10^{-13}f_{\bar U} (S_6)
\label{rUfr}
\end{eqnarray}
at $T \sim T_{fU} \approx m_U/30 \approx 100 GeV$. Here $f_{U} (1)=f_{\bar U} (1)=1$ and their functional form is given in Appendix 1. This functional form is simplified for large $S_6 > 1$
\begin{eqnarray}
r_U  \approx 8 \cdot 10^{-13}S_6\cdot (1 - \ln{(S_6)}/30) + 6 \cdot 10^{-14}/S_6\nonumber\\
r_{\bar U}  \approx  8 \cdot 10^{-13}S_6\cdot (1 - \ln{(S_6)}/30) - 6 \cdot 10^{-14}/S_6
\label{rUSbpm}
\end{eqnarray}
and for smallest possible $0.2 < S_6 < 0.4$
\begin{eqnarray}
r_U  \approx \kappa_U = 1.2 \cdot 10^{-13}/S_6\nonumber\\
r_{\bar U}  \approx 1.1 \cdot 10^{-14} \exp\left( -0.16/S_6^2 \right)
\label{rUSspm}
\end{eqnarray}

It means that the concentration of frozen out  $U$-quark pairs is
for $S_6=1$ by 6 times larger than the concentration of excessive
$U$-hadrons Eq.(\ref{Uexcess}) and this effect grows with $S_6$ as
$\propto S_6^2$ at large $S_6$. Some suppression of $\bar U$-quark
abundance takes place only for smallest possible values of $S_6$,
but even in this case it can not be less than $r_{\bar U}  \approx
2 \cdot 10^{-16},$ which is reached at $S_6 = 0.2.$ So in this
moment, in spite of assumed tera-baryon asymmetry, the frozen out
concentration of antiquarks $\bar U$ is not strongly suppressed
and they can not be neglected in the cosmological evolution of
tera-fermions.

\subsection{\label{Efreezing} Freezing out of $E$-leptons}
The same problem of  antiparticle survival
appears (enhanced) for $E$-leptons.
Equilibrium concentration of $E E^+$ pairs starts to decrease at $T<m_E=500 GeV S_6$.
At the freezing out temperature $T_f$ the rate of expansion
exceeds the rate of annihilation to photons $E E^+ \rightarrow
\gamma \gamma$ or to pairs of light fermions $f$ (quarks and charged leptons) $E E^+ \rightarrow
\bar f f$ (We neglect effects of $SU(2)$ mediated bosons). Then $E$ leptons and their antiparticles $E^+$ are frozen out.

The frozen out
concentration (in units of entropy density) of $E$, $r_{E}$, and $E^+$, $r_{E^+}$,
is given  (see Appendix 1) by
\begin{eqnarray}
r_E= 10^{-11} S_6\cdot (1 - \ln{(S_6)}/25)+0.4 \cdot 10^{-13}/S_6 \nonumber\\
r_{E^+}= 10^{-11} S_6\cdot (1 - \ln{(S_6)}/25)- 0.4 \cdot 10^{-13}/S_6
\label{rEfr}
\end{eqnarray}
at $T \sim T_{fE} \approx m_E/25 \approx 20 GeV S_6$.
One finds from Eq.(\ref{rEfr}) that at $S_6=1$ the frozen out concentration
of $E E^+$ pairs is by 2 orders of magnitude larger
than the concentration Eq.(\ref{Eexcess}) of excessive E and this effect increases $\propto S_6^2$ for larger and larger $S_6$. Even at smallest possible $S_6$ $E E^+$ pair abundance is 5 times larger than $L'$ excess.

Antiparticles $\bar U$ and $E^+$ should be effectively annihilated in the successive processes
of quark and $E$ recombinations. However, as it is shown in Appendices 3-5 primordial anti-quark tera-hadrons can be effectively suppressed, while as we'll see similar mechanism of annihilation is not effective for tera-positrons.

\subsection{\label{anE} $E^+$ annihilation in twin tera-positronium $E E^+$}

The frozen out $E^+$
can bind at $T< I_{E^+}= \alpha^2 m_E/4 \approx 6.3 MeV S_6$ with $E$ into positronium-like systems and annihilate. Since the lifetime of these positronium-like systems is much less, than the timescale of their disruption by energetic photons, the direct channel of $E^+$ binding in $E E^+$ and annihilation can not be compensated by inverse reaction of photo-ionization.
That is why, similar to the case of chromo-Coulomb binding  of $\bar U$ and their annihilation in "$U$-charmonium", considered in Appendix 3, $E^+$ begin to bind with $E$ and annihilate as soon as temperature becomes less than $I_{E^+}= \alpha^2 m_E/4 \approx 6.3 MeV S_6$.
The decrease of $E^+$ abundance owing to $E E^+$
recombination is governed by the equation
\beq
\frac{dr_{E^+}}{dt} = -r_E r_{E^+} \cdot s \cdot \sv,
\label{hadrecombE}
\eeq
where $s$ is the entropy density and (see Appendix 2)
$$\sv=  (\frac{16\pi}{3^{3/2}}) \cdot \frac{ \alpha}{T^{1/2}\cdot m_E^{3/2}}.$$
Using the formalism of Appendix 1 we can re-write the Eq.(\ref{hadrecombE}) as
\beq
\frac{dr_{E^+}}{dx}=f_{1E^+}\left<\sigma v\right> r_{E^+}(r_{E^+}+\kappa_E),
\label{Esrecomb}
\eeq
where $x=T/I_{E^+}$, the asymmetry $\kappa_E =  r_E-r_{E^+} =r_{L'} =8 \cdot 10^{-14}/S_6$ is given by Eq.(\ref{Eexcess}) and
$$f_{1E^+}=\sqrt{\frac{\pi g_s^2}{45g_{\epsilon}}}m_{Pl}I_{E^+} \approx m_{Pl}I_{E^+}.$$
The concentration of remaining $E^+$ is given by
\beq
r_{E^+}=\frac{\kappa_E\cdot r_{fE^+}}{(\kappa_E+r_{fE^+}) \exp\left( \kappa_E J_{E^+} \right) - r_{fE^+}},
\label{rEbar}
\eeq
where from Eq.(\ref{rEfr})
$$r_{E^+}= 10^{-11} S_6\cdot (1 - \ln{(S_6)}/25)- 0.4 \cdot 10^{-13}/S_6$$
and
$$J_{E^+}=\int_0^{x_{fE^+}} f_{1E^+}\left<\sigma v\right>dx =$$
\beq
= m_{Pl}I_{E^+} 4 \pi (\frac{2}{3^{3/2}}) \cdot \frac{ \alpha^2}{I_{E^+}\cdot m_E} \cdot 2 \cdot x_{fE^+}^{1/2}\approx
1 \cdot 10^{13}/S_6.
\label{JEbar}
\eeq
In the evaluation of Eq.(\ref{JEbar}) we took into account that the decrease of $E^+$ starts at $T\sim I_{E^+}$, so that $x_{fE^+} \sim 1$.
In the case of $E^+$ the reaction rate $\sv$ in Eq.(\ref{JEbar}) contains square of fine structure constant instead of square of QCD constant $\bar \alpha$ in the case of $\bar U$. It makes the situation with $E^+$ at $S_6 \sim 1$ principally different from the case of antiquarks: the abundance of $\bar U$ is suppressed exponentially, while for $E^+$ exponential suppression is practically absent.
Indeed, one has $\kappa_E J_{E^+} \approx 0.8/S_6^2$ in the exponent of Eq.(\ref{rEbar}).
For all $S_6$ the condition $r_{fE^+} \gg \kappa_E$ is valid. Therefore the solution Eq.(\ref{rEbar})
has the form
\beq
r_{E^+} \approx \frac{\kappa_E}{ \exp\left( \kappa_E J_{E^+} \right) - 1},
\label{rEbars}
\eeq
which gives for $S_6 > 1$
$$r_{E^+} \approx \frac{1}{ J_{E^+} }  -  \frac{\kappa_E}{ 2 } \approx 1.4 \cdot 10^{-13} S_6 - 0.4 \cdot 10^{-13}/S_6.$$
In the result the residual amount of $E^+$ remains at $S_6 \ge 1$ enormously high, being  for  $S_6 \sim 1$  larger than $L'$ excess.

At smallest allowed values of $S_6<1$ $E^+$ abundance is suppressed
$$r_{E^+}= \kappa_E \cdot  \exp\left(- \kappa_E J_{E^+} \right) \approx \left(\frac{8\cdot 10^{-14}}{S_6}\right) \exp\left( -0.8/S_6^2\right)$$
and for the minimal value $S_6=0.2$ the abundance of primordial tera-positrons falls down to
$$r_{E^+} \approx 4 \cdot 10^{-13}\exp\left( -20\right)  \approx 1.6 \cdot 10^{-21}.$$
Even so suppressed a light Sinister Universe still provides a huge tera-lepton over-abundance
(see section 5).
On the other hand, this lowest tera-positron abundance lets the tera electrons amount at small values of $S_6<1$ close to the asymmetric excess $\kappa_E =  r_E-r_{E^+} =r_{L'} =8 \cdot 10^{-14}/S_6.$

The general expression for tera electron abundance $r_E$ after twin tera-positronium annihilation has the form (see Appendix 1)
$$r_E=\frac{\kappa_E\cdot r_{Ef}}{r_{Ef}-(r_{Ef}-\kappa_E) \exp\left(-\kappa_E J_{E^+}\right)},$$
where $J_{E^+}$ is given by Eq.(\ref{JEbar}) and from Eq.(\ref{rEfr})
$$r_E= 10^{-11} S_6\cdot (1 - \ln{(S_6)}/25)+0.4 \cdot 10^{-13}/S_6.$$
With the account for $r_{Ef} \gg \kappa_E$ for all $S_6$ one obtains
\beq
r_E=\frac{\kappa_E}{1-\exp\left(-\kappa_E J_{E^+}\right)}.
\label{rEtpan}
\eeq
It tends to $r_{E} \approx 1/J_{E^+}  + \kappa_E/2 \approx 1.4 \cdot 10^{-13} S_6 + 0.4 \cdot 10^{-13}/S_6$ at large $S_6$ and to $\kappa_E$ for small $S_6 < 1$.

\subsection{\label{recE} E-Ubaryon recombination}
At the temperature $T< I_{UE} =Z^2 \alpha^2 m_E/2 \approx 50 MeV S_6$ (where electric charge of $(UUU)$ is $Z=2$) $(UUU)$ can form atom-like systems with $E$. Reactions
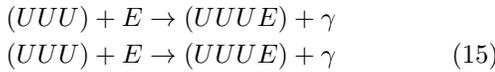
\begin{eqnarray}
(UUU) + E \rightarrow (UUUE) + \gamma\nonumber\\
(UUU) + E \rightarrow (UUUE) + \gamma
\label{EUUUbind}
\end{eqnarray}
are balanced by inverse reactions of photo-destruction. According to Saha-type equations effective formation of $(UUUE)$ and $(UUUEE)$ systems is delayed until $T_{fUE} \sim I_{UE}/30 \sim 1.5MeV S_6.$

In this period composite $(UUUEE)$ cold dark matter is formed. However, though most of $(UUU)$ bind with $EE$ into $(UUUEE)$, significant fraction of free $(UUU)$ and $(UUUE)$ remains unbound, what we show below.

In the considered period $r_{UUU} = r_{B'}=4 \cdot 10^{-14}/S_6$ (In the following we assume sufficiently effective suppression of $\bar U$ hadrons in hadronic recombination, as it is in cases A and B of Appendix 5), while the abundance of $E$ after incomplete annihilation with $E^+$ is $r_{E} = 1.4 \cdot 10^{-13} S_6 + 0.4 \cdot 10^{-13}/S_6$ at $S_6 \ge 1$ and only at  $S_6 < 1$ tends to $\kappa_E = 8 \cdot 10^{-14}/S_6$. At $T<T_{fUE} \sim I_{UE}/30 \sim 1.5MeV S_6$ the residual amount of  free $(UUU)$ is governed by equation
\beq
\frac{dr_{UUU}}{dx}=f_{1EU}\left<\sigma v\right> r_{UUU}(r_{UUU}+\kappa_{UE}),
\label{EUUUrecomb}
\eeq
where $x=T/I_{UE}$ and $\kappa_{UE} =  r_E-r_{B'}.$ At $S_6=1$ $\kappa_{UE} =1.4 \cdot 10^{-13}$, at large $S_6 \gg 1$ $\kappa_{UE} =1.4 \cdot 10^{-13} S_6 $, while at smallest $S_6 \sim 0.2$  the value of $\kappa_{UE}$ is approximately $4 \cdot 10^{-14}/S_6.$
In the Eq.(\ref{EUUUrecomb})
$$\sv=  (\frac{4\pi}{3^{3/2}}) \cdot \frac{ \alpha^2 Z^2}{I_{UE}\cdot m_E}\frac{1}{x^{1/2}}$$
and
$$f_{1EU}=\sqrt{\frac{\pi g_s^2}{45g_{\epsilon}}}m_{Pl}I_{UE} \approx m_{Pl}I_{UE}= m_{Pl}Z^2 \alpha^2 m_{E}/2.$$
Solution of Eq.(\ref{EUUUrecomb}) is given by
\beq
r_{UUU}=\frac{\kappa_{UE}\cdot r_{UUUf}}{(\kappa_{UE}+r_{UUUf}) \exp\left(\kappa_{UE} J_{UE} \right) - r_{UUUf}}.
\label{rEUUU}
\eeq
Here $r_{UUUf}<r_{B'}$ is the abundance of $(UUU)$ at $T_{fUE} \sim I_{UE}/30$,
$$J_{UE}=\int_0^{x_{fUE}} f_{1EU}\left<\sigma v\right>dx =$$
\beq
= m_{Pl} (\frac{2 \pi}{3^{3/2}}) \cdot \frac{Z^2 \alpha^2}{m_E} \cdot 2 \cdot \sqrt{x_{fUE}}\approx
4 \cdot 10^{12}/S_6
\label{JEUUU}
\eeq
and we took ${x_{fUE}} \sim 1/30$. For $S_6\ge 1$ in the exponent of solution Eq.(\ref{rEUUU}) $\kappa_{UE} J_{UE} \approx 0.56$ is independent of $S_6$. Therefore the following approximate expression is valid for $S_6 \ge 1$
\beq
r_{UUU}=\frac{r_{UUUf}}{1+(\kappa_{UE}+r_{UUUf})J_{UE}} \approx r_{UUUf}.
\label{rEUUUapr}
\eeq
Similar arguments are valid for $U$- "ions" ($(UUu)^{++}$, $(UUuE)^+$, $(UUUE)^+$ etc).

At small $S_6<0.4$ the value of $\kappa_{UE} J_{UE} \approx 0.16/S_6^2$ exceeds 1, being equal to 4 at $S_6=0.2.$ It does not involve principal changes in our conclusions.

The abundance of $(UUU)$ and $(UUUE)$, which remain free, is determined by its value at $T\sim T_{fUE}$. The ratio $\frac{n_{(UUU)}}{n_{(UUUE)}} \sim \frac{n_{(UUUE)}}{n_{(UUUEE)}}$ can hardly be much less, than  $0.1$. The same is true for all the other $U$-ions.

Binding of $E$ with $(UUU)$ and $(UUUE)$ decreases the abundance of $E$, but to the end of the first second of cosmological expansion the relic tera-lepton pairs of $E$ and $E^+$ still remain the dominant form of tera-matter.

In the successive analysis we assume for definiteness at $S_6>1$
\beq
\frac{n_{(UUU)}}{n_{(UUUE)}} \sim \frac{n_{(UUUE)}}{n_{(UUUEE)}} \sim \frac{1}{10}
\label{recfrUUUE}
\eeq
with the same proportion for $(UUu):(UUuE):(UUuEE)$
and for $(Uuu):(UuuE):(UuuEE)$.
If the lightest $U$-hadron is $(Uud)$, we assume
\beq
\frac{n_{(Uud)}}{n_{(UudE)}} \sim \frac{1}{10}.
\label{recfrUudE}
\eeq
For smallest $S_6 \sim 0.2$ the above proportions may be an order of magnitude smaller.

\subsection{\label{briefSum} Brief summary of Sinister trace at $t \sim 1\s$}
Under these assumptions the tera-matter content of the Universe to the end of MeV era is:

1. Free $E$ with $r_{E}=1.08 \cdot 10^{-13}$ at $S_6 \sim 1$. It is $(1.4\cdot S_6  - 0.32/S_6) \cdot 10^{-13}$ at $S_6 \gg 1$ and tends to $4 \cdot 10^{-16}/S_6$ at $S_6 \rightarrow 0.2.$

2. Free $E^+$ with $r_{E^+}=1 \cdot 10^{-13}$ at $S_6 \sim 1$, growing to $(1.4\cdot S_6  - 0.4/S_6) \cdot 10^{-13}$ for $S_6 \gg 1$ and decreasing down to $1.6 \cdot 10^{-21}$ at $S_6 = 0.2.$

3. Neutral $(UUUEE)$ "tera-helium-atoms" (with $r_{(UUUEE)} \approx 3.6 \cdot 10^{-14}/S_6$
for $S_6 \ge 1$, growing up to $3.96 \cdot 10^{-14}/S_6$, when $S_6$ decreases down to 0.2)
with an uncertain admixture (up to $10\%$) of first and second hybrid tera-helium atoms $(UUuEE)$ and $(UuuEE)$. The minimal estimation for this admixture is $r_{(UUuEE)} \sim r_{(UuuEE)} \sim 10^{-20}$. If $(Uud)$ is the lightest $U$-hadron, there should be neutral hybrid tera-hydrogen atoms with minimal abundance $r_{(UudE)} \sim 10^{-20}$.

4. Charged $(UUUE)^+$ "tera-helium-I-ion" with $r_{(UUUE)} \approx 4 \cdot 10^{-15}/S_6$
for $S_6 \ge 1$ and decreasing down to $r_{(UUUE)} \approx 4 \cdot 10^{-16}/S_6$ at $S_6 \approx 0.2$.

5. Free double charged $(UUU)$ pure tera-helium ion with
$r_{(UUU)} \approx 4 \cdot 10^{-16}/S_6$ for $S_6 \ge 1$ and tending to $r_{(UUU)} \approx 4 \cdot 10^{-18}/S_6$ at $S_6 \rightarrow 0.2$.

6. Theoretically uncertain amount of double charged "first hybrid tera-helium ions"  $(UUu)$ - relics of hadronic recombination with the abundance in the range  $10^{-20} \le r_{(UUu)} \le 4 \cdot 10^{-15}/S_6$. We'll take for definiteness the minimal estimation $r_{(UUu)} =10^{-20}.$

7. Theoretically uncertain amount of theoretically uncertain lightest "tera-baryon" $(Uqq)$ ($(Uuu)$ or $(Uud)$) $10^{-20} \le r_{(Uqq)} \le 4 \cdot 10^{-15}/S_6$. The minimal estimation $r_{(Uqq)}= 10^{-20}$ will be taken for definiteness.

8. Exponentially small amount of stable tera-mesons $(\bar U u)$ (We accept case A in hadronic recombination, what results in their suppression $\propto \exp{(-4 \cdot 10^{5}/S_6)}).$


For $S_6 \ge 1$ pairs of free tera-leptons $E$ and $E^+$ dominate among these relics from MeV era. The abundance of these pairs relative to $L'$ excess grows with $S_6 >1$ as $\propto S_6^2$. Since mass of tera-lepton is $\propto S_6$, the contribution of tera-lepton pairs into total density grows as $\propto S_6^3$ relative to $L'$ excess. If survive, tera-lepton pairs can over-close the Universe even at modest $S_6 >1$ and such survival seems inevitable on the following reason.
In Big Bang Nucleosynthesis $^4$He is formed with abundance $r_{He} = 0.1 r_b$ and due to larger binding energy it can bind with tera-electrons earlier, than $p$. Instead of neutral $(Ep)$ atom tera-electrons form positively charged $(^4HeE^-)^+$ ion. Coulomb barrier makes impossible reactions of $E^-$, trapped in this ion, with other positively charged tera-remnants ($E^+$ and $U$-ions). On the other hand virtually all the free $E^-$ are captured  by $^4$He before $Ep$ binding is possible and there are no free $E^-$, which can form $(E^-p)$.
It leaves no hope to suppress the above list of tera-matter remnants with the use of $(Ep)$
catalysis  \cite{Glashow}.

However, in the process of SBBN reactions and its binding with tera-electron can influence the SBBN reactions, as well as it strongly changes the picture of successive evolution of tera-matter, leaving no room for the hope \cite{Glashow} to suppress the above list of tera-matter remnants with the use of $(Ep)$ catalysis.
\section{\label{catEHe} Helium-4 cage for free $E^-$}
At $T<I_{EHe} = Z^2 \alpha^2 m_{He}/2 \approx 400 keV$ reaction
\beq
E+^4He\rightarrow \gamma +(^4HeE)^+
\label{EHeIg}
\eeq
can take place. In the successive reaction
\beq
E+(^4HeE)^+\rightarrow \gamma +(^4HeEE)
\label{EHeg}
\eeq
tera-helium $(EEHe)$ "atom" is produced. The size of this "ion" and "atom" is
$$R_{EHe} \sim 1/(Z \alpha m_{He}) \approx 4 \cdot 10^{-13} cm$$
and they can play nontrivial catalyzing role in the nuclear transformations of SBBN.

For our problem another aspect is important.  Reactions Eqs.(\ref{EHeIg}) and  (\ref{EHeg})
can start only after $^4$He is formed, what happens at $T<100 keV$. Then inverse reactions of ionization by thermal photons support Saha-type relationships between the abundances of these "ions", "atoms", free $E^-$, $^4$He and $\gamma$:
\beq
\frac{n_{He} n_E}{n_{\gamma} n_{(EHe)}} = \exp{(-\frac{I_{EHe}}{T})}.
\label{recHeSaha1}
\eeq
and
\beq
\frac{n_E n_{(EHe)}}{n_{\gamma} n_{(EEHe)}} = \exp{(-\frac{I_{EHe}}{T})}.
\label{recHeSaha2}
\eeq
When $T$ falls down below $T_{rHe} \sim I_{EHe}/\log{\left(n_{\gamma}/n_{He}\right)} \approx I_{EHe}/27 \approx 15 keV$ free $E^-$ are effectively bound with helium in reaction Eq.(\ref{EHeIg}).
The fraction, which
forms neutral $(^4HeE^-E^-)$ depends on the ratio of $E^-$ and $^4$He abundances. For $S_6<57$ this ratio is less, than 1. Therefore, when, owing to $^4$He excess, virtually all $E^-$ form $(^4HeE^-)^+$ ion in reaction  Eq.(\ref{EHeIg}), there are no free $E^-$ left to continue binding in reaction Eq.(\ref{EHeg}). Moreover, as soon as neutral $(^4HeEE)$ is formed it catalyzes reactions of $UE$ binding
\beq
UUU+(EEHe) \rightarrow (UUUEE) +^4He
\label{EHeUUU}
\eeq
\beq
(UUUE) + (EEHe) \rightarrow (UUUEE)+^4He +E
\label{EHeUUUE}
\eeq
as well as tera-positron annihilation through twin tera-positronium formation
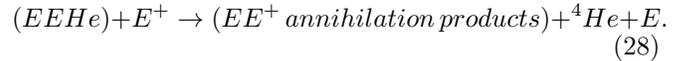
\beq
(EEHe)+E^+ \rightarrow (EE^+\,annihilation\,products ) +^4He +E.
\label{EHeP}
\eeq
In these reactions heavy $U$-ion or tera-positron penetrates neutral $(EEHe)$ "atom" and
expel $^4He$.  $U$-ions form terahelium "atom". Tera-positron forms twin tera-positronium ion $(EEE^+)^-$ with charge $Z=-1$. In the latter one of $E^-$ annihilates with $E^+$, leaving free $E^-$.
\subsection{\label{BackP} $(^4HeE^-)$ trap surviving back reaction of $E^+$ annihilation}
Energetic particles, created in $EE^+$ annihilation, interact with cosmological plasma.
In the development of electromagnetic cascade creation of electron-positron pairs in the reaction $\gamma + \gamma \rightarrow e^+ + e^-$ plays important role in astrophysical conditions (see \cite{BL,AV,book} for review). The threshold of this reaction puts upper limit on the energy on the nonequilibrium photon spectrum in cascade
\beq
E_{max} = a\frac{m_{e}^2}{25T},
\label{Emax}
\eeq
where factor $a = \ln{(15 \Omega_b + 1)} \approx 0.5$.

At $T>T_{rbHe} = a m^2_e/(25 I_{He}) \approx 12.5 keV$ in the spectrum of electromagnetic cascade from $EE^+$ annihilation maximal energy $E_{max} <I_{He}$
and $E^+$ annihilation products can not ionize $(^4HeE^-)$ and $(^4HeE^-E^-)$. So, there is no back reaction of $E^+$ annihilation until $T \sim T_{rbHe}$ and in this period practically all free $E^-$ are bound in $(^4HeE^-)^+$ ion. Due to Coulomb barrier $E^+$ can not penetrate $(^4HeE^-)^+$ ion and annihilate with $E^-$ in it.

Small fraction ($\sim r_{E}^2/r_{He}$) of $(^4HeEE)$ is initially formed in the case $S_6 < 57$ but immediately eaten by tera-positrons. It leads to corresponding small decrease of initial tera positron abundance and of abundance of $(^4HeE^-)^+$ relative to initial amount of $E^-$.
This decrease of $r_{E^+}$ is $\sim 1/2$ for smallest possible value $S_6=0.2$.
Therefore virtually all primordial tera-leptons remain in the Universe and contribute into its total density. For $S_6 > 1.7$ this contribution exceeds $\Omega_{CDM}$.

The case $S_6>57$, when $r_E > r_{He}$, is even more troublesome, since in this case all the $^4$He, produced in SBBN, is bound in $(^4HeE^-)^+$. The successive formation of neutral $(^4HeE^-E^-)$
is now possible, so all the excessive tera-positrons $r_{E^+} > r_{He}$ annihilate with $E^-$ in $(^4HeE^-E^-)$ until their abundance decreases down to $r_{E^+} = r_{He}$, when they eliminate all the $(^4HeE^-E^-)$. In the result $r_{E^+} \sim r_{He}$ and $r_{(E^4He)^+} \sim r_{He}$ are left, leading to huge over-closure of the Universe ($\Omega > 5 \cdot 10^3 \Omega_{CDM}$).

\section{\label{Matter} The $Sinister$ overproduction of Anomalous Hydrogen clones}
The main problem of the considered cosmological scenario is the over-production of primordial tera-lepton pairs and their conservation in the Universe in various forms up to present time.

In the period of recombination of nuclei with ordinary electrons ($e$),
$(^4HeE^-)^+$, $E^+$, free charged $U$-baryons,
as well as charged $(UUUE)$, $(UUuE)$, $(UuuE)$ bound systems
recombine with electrons to form
atoms of anomalous isotopes. The substantial (no less than 6 orders
of magnitude) excess of electron number density over the number density
of primordial tera-fermions makes virtually all of them to form atoms (see Appendix 6).

At $S_6 > 1$ contribution of these atoms in the total density exceeds $\Omega_{CDM}$. Therefore only a small interval $0.2 < S_6 < 1$ can be considered. Then the dominant form of tera matter is neutral $(UUUEE)$, which saturates the total Dark matter density and might drive the development of gravitational instability, resulting in galaxy formation. This neutral tera matter contains an uncertain fraction of hadronic tera-helium $(UUuEE)$ and $(UuuEE)$ (or tera-hydrogen $(UudE)$, if $(Uud)$ is the lightest). Though the total contribution of this fraction to the DM density  should be less than $10\%$ it can not be less than $2.5 \cdot 10^{-7}.$
Moreover, the dominant form of tera-matter is accompanied by other forms of tera-matter with the following abundances for $S_6 = 1$ (we also give lower limit at $S_6 = 0.2$)
$$\xi_i = r_{i}/r_b$$
relative to baryons:

1. Hybrid tera-helium $(^4HeE^-e)$ with $\xi_{(eEHe)}=r_{(eEHe)}/r_b \approx 1.4 \cdot 10^{-3}$,
(at $S_6= 0.2$  $\xi_{(eEHe)} \approx 2.5 \cdot 10^{-5}$)

2. Hybrid tera-positronium $(eE^+)$ with $\xi_{(eE^+)} \approx 1.3 \cdot 10^{-3}$,
(at $S_6= 0.2$  $\xi_{(eE^+)} \approx 1 \cdot 10^{-11}$)

3. Tera-helium I hybrid atoms $(UUUEe)$ with $\xi_{(UUUEe)} \approx 5 \cdot 10^{-5}$,
(at $S_6= 0.2$  $\xi_{(UUUEe)} \approx 5 \cdot 10^{-6}$)

4. Pure tera-helium hybrid atoms $(UUUee)$ with $\xi_{(UUUee)} \approx 5 \cdot 10^{-6}$
(at $S_6= 0.2$  $\xi_{(UUUee)} \approx 2.5 \cdot 10^{-7}$)

5. Pure, first and second tera-helium I atoms $(UUUEe)$, $(UUuEe)$, $(UuuEe)$ with abundance, no less than $\xi_{eU_i} \approx 1.3 \cdot 10^{-10}$ (at $S_6= 0.2$  $\xi_{(eU_i)} \approx 1.3 \cdot 10^{-11}$)

6. First and second hybrid tera-helium atoms $(UUuee)$ and $(Uuuee)$ with abundance, no less than $\xi_{eeU_i} \approx 1.3 \cdot 10^{-11}$ (at $S_6= 0.2$  $\xi_{(eeU_i)} \approx 5 \cdot 10^{-13}$)

and exponentially small amount of free $E$, $(UUU)$, $(UUu)$, $(Uuu)$ and $\bar U$ hadrons.

All these $e$-atoms, having atomic cross sections of interaction with
matter, participate then in formation of astrophysical bodies,
when galaxies are formed. At $S_6 = 1$ density
of anomalous hydrogen $(^4HeE^-e^-)$, $(eE^+)$ and $(UUUEe)$ is
two times larger, than baryonic density. For all the range $0.2 < S_6 <1$ their abundance is
many orders of magnitude higher than the experimental upper limits \cite{exp1,exp2,exp3,exph1,exph2,exph3}. Moreover there does not seem to be any mechanism to reduce their primordial abundance in matter bodies.







To save the sinister model from this trouble the following mechanisms were mentioned in \cite{Glashow}:

- more complete aggregation of remnants into tera-helium during the structure formation

- gravitational concentration of heavy remnants inside stars

- procession of remnants into superheavy elements other than those for which sensitive searches were carried out.

Both the first and the last mechanisms are not appropriate for leptonic anomalous hydrogen $eE^+$
and it may be shown that for all the types of remnants gravitational concentration in stars is not effective (Appendix 7).

Moreover, the mechanisms of the above mentioned kind can not in principle suppress the abundance of remnants in interstellar gas more than by factor $f_g \sim 10^{-2}$, since at least $1\%$ of this gas has never passed through stars or dense regions, in which such mechanisms are viable. It may lead to the presence of $(^4HeE^-)^+$, $E^+$, $(UUUE)$ and other fragment's component of cosmic rays at a level $\sim f_g \xi_i$. Therefore based on sinister model with $1 \ge S_6 \ge 0.2$ one can expect the anomalous hydrogen fractions of cosmic rays
\begin{eqnarray}
10^{-5} \ge \frac{(^4HeE^-)^{+}}{p} \ge 2.5 \cdot 10^{-7},\nonumber\\
10^{-5} \ge \frac{E^{+}}{p} \ge 10^{-13},\nonumber\\
5 \cdot 10^{-7} \ge \frac{(UUUE)}{p} \ge 5 \cdot 10^{-8}
\end{eqnarray}
and anomalous helium
$$5 \cdot 10^{-7} \ge \frac{(UUU)}{He} \ge 2.5 \cdot 10^{-8}.$$
If $(^4HeE^-)^+$ is disrupted  in the course of cosmic ray acceleration one should expect anomalous charge $Z=-1$ component of cosmic rays
$$10^{-5} \ge \frac{E^-}{p} \ge 2.5 \cdot 10^{-7}.$$

These predictions may be within the reach for future cosmic ray experiments, in particular, for AMS.

The only way to solve the problem of anomalous isotopes is to find a possible reason for their low abundance inside the Earth and solution of this problem implies a mechanism of effective suppression of anomalous hydrogen in dense matter bodies
(in particular, in Earth).

The idea of such suppression, first proposed in \cite{fractons} was recently realized in \cite{4had}. However we don't have this possibility for tera-matter, unless the underlying model is modified to include long range attraction for tera-particles. Another possible modification of the underlying sinister model is to provide the mechanism of tera-particle instability.
If their lifetime is less, than the age of Universe, they decay and there is no problem of their over-abundance in the present Universe, but the effects of the decay products should satisfy astrophysical constraints (see review and Refs in \cite{book}). In accelerator search, however, unstable particles with lifetime, much less than the age of Universe can be treated as stable (see Appendix 8).

\section{\label{discussion} Discussion}

Experimental data on $Z$ boson width exclude the possibility for new heavy generations,
 constructed in the same way as the three known families.
 New heavy quarks and leptons should possess some new property,
 which differs them from fermions of known generations.

Such property may be a new long range interaction, which new (fourth)
 generation possesses \cite{4had}. With the use of this interaction an effective mechanism,
 reducing the abundance of anomalous isotopes below the existing upper limit can be realized \cite{fractons,4had}.
  The dominant form of dark matter can not be explained in this framework.

The idea of novel Glashow's sinister model for heavy fermions is
more ambitious, since together with generation of neutrino mass
and solution of strong CP violation problem it pretended to
explain the dominant form of dark matter by bound $(UUUEE)$
tera-helium "atoms". The nontrivial possibility to reproduce CDM
scenario and may be to resolve the puzzle of controversial results
of direct WIMP searches is inevitably accompanied in this model by
prediction of various forms of anomalous isotopes and, first of
all, of anomalous hydrogen $(eE^4He)$, $(eE^+)$ and $(UUUEe)$.
Overproduction of these exotic forms of tera-matter is a serious
problem for the considered scenario.
Having fixed the $B'-L'$ asymmetry and mass of tera-particles
 we can not avoid a substantial amount of primordial frozen out $E E^+$ pairs.
 This amount grows relative to $(B'-L')$-asymmetric excess $\propto S_6^3$ and,
 if these pairs are not annihilated, at $S_6 > 1.7$ exceeds $\Omega_{CDM}$.
  the other side of tera-lepton catastrophe is trapping of free $E^-$ in $(^4HeE^-)^+$,
  which inhibits such annihilation and precludes effective decrease of tera-lepton primordial abundance.
   Even for minimal value $S_6 = 0.2$ the predicted terrestrial abundance of anomalous hydrogen exceeds experimental
   upper limits by more than 20 orders of magnitude.


The problem of  primordial tera-lepton
overproduction can not be resolved for the present version
  of sinister model. Some additional physics is needed to provide effective mechanism
  of tera-lepton suppression. It might imply the necessity for tera-particles
   to be unstable with the lifetime less than the age of the Universe.
   Then they still can be considered as stable in accelerator experiments.
   Meta-stable $m > 100 GeV$  $E$ and $E^+$ and single $U$-quark hadrons become
   in this case the challenge for new generation of particle accelerators.

If survives, sinister Universe offers new interesting framework
for particle physics and cosmology. If not, there still can be
some place for tera-matter with its flavor of possible composite
dark matter in Galaxy and exotic rare forms of stable matter
around us.

\section{\label{conclusion} Conclusions}

In conclusion, tera-leptons hidden in $(^4HeE^-)^+$ cage or frozen
in hybrid tera-positronium $(eE^+)$ as well as in other hybrid
components are guaranteed lethal relics for any Sinister Universe.
Behaving as anomalous hydrogen isotope most of these relics suffer
of all the correlated and severe bounds on our lightest elements.

The $S_6$ parameter does not offer any way out or escape,
 wherever it grows or decreases: the tera-leptons are ever offending atomic data records.
  While exotic and {\it ad hoc} tera-lepton interaction might offer a hope to survive,
  they are nevertheless breaking the Sinister beautiful simplicity and symmetry.
   There is in the unstable scenario for tera-Leptons and Tera-Quarks a very interesting case for UHECR
    as discussed in Appendix 9, able in principle to
    offer a solution to GZK puzzle.

We believe that the present study and collapse of our toy model
will offer a clarifying starting point to new and more lucky
scenarios, where Heavy Lepton and Heavy Quark may better fit the
nature puzzles.

  In some sense we feel that while with regret we are closing the door to a
  Glashow's Sinister Cosmology we are disclosing a fascinating windows for
  meta-stable tera-particle  physics and  astrophysics.

\section*{\label{Ackn} Acknowledgment}
We  are very grateful to S.L. Glashow for inspiring with his
paper, his letters and support the evolution of our study; we
thank K.Belotsky and M.Ryskin for fruitful discussions and help.
 M.Kh. is grateful to CRTBT-CNR, Grenoble and LPSC, Grenoble for
kind hospitality. \small



\section*{Appendix 1. Charge asymmetry in freezing out of particles and antiparticles
} The frozen number density of cosmic relics, which were in
equilibrium with primordial plasma, is conventionally deduced
\footnote{We are grateful to K.Belotsky for help in preparation of
this Appendix.} from equation \cite{ZeldNov}
\begin{equation}
\dot{n}+3Hn=\left<\sigma_{ann}v\right>(n_{eq}^2-n^2).
\label{sym}
\end{equation}
This equation is written for the case of a charge symmetry of particles in question, i.e. for the case
when number densities of particles ($X$) and antiparticles ($\bar{X}$) are equal $n_X=n_{\bar{X}}=n$.
The value $n_{eq}$ corresponds to their equilibrium number density
and is given by Boltzmann distribution
\begin{equation}
n_{eq}=g_S \frac{mT}{2\pi}^{3/2}\exp \left(-\frac{m}{T}\right).
\label{neq}
\end{equation}
Here $g_S$ and $m$ are the number of spin states and the mass of given particle.

In course of cooling down $n_{eq}$ decreases exponentially and
becomes below freezing out temperature $T_f$ much less then real density $n$,
so the term $\left<\sigma_{ann}v\right>n_{eq}^2$, describing creation of $X\bar{X}$ from
plasma, can be neglected \cite{Turner}.
It allows to obtain approximate solution of Eq.(\ref{sym}).

In case of charge asymmetry one needs to split Eq.(\ref{sym}) on two: for $n_X$ and $n_{\bar{X}}$,
which are not equal now.
\begin{eqnarray}
\dot{n}_X+3Hn_X=\left<\sigma_{ann}v\right>(n_{eq\,X}n_{eq\,\bar{X}}-n_{X}n_{\bar{X}}),\nonumber\\
\dot{n}_{\bar{X}}+3Hn_{\bar{X}}=\left<\sigma_{ann}v\right>(n_{eq\,X}n_{eq\,\bar{X}}-n_{X}n_{\bar{X}}).
\label{asym}
\end{eqnarray}
The values $n_{eq\,X}$ and $n_{eq\,\bar{X}}$ are given by Eq.(\ref{neq}) with inclusion of chemical potential,
which for $X$ and which for $\bar{X}$ are related as $\mu_X=-\mu_{\bar{X}}=\mu$ (see, e.g., \cite{Dolgov}). So
\begin{equation}
n_{eq\,X,\bar{X}}=\exp\left(\pm\frac{\mu}{T}\right) n_{eq},
\label{nmueq}
\end{equation}
where upper and lower signs are for $X$ and $\bar{X}$ respectively. So
\begin{equation}
n_{eq\,X}n_{eq\,\bar{X}}= n_{eq}^2.
\label{neq2}
\end{equation}

A degree of asymmetry will be described in conventional manner (as for baryons)
by the ratio of difference between $n_{X}$ and $n_{\bar{X}}$
to number density of relic photons at the modern period
\begin{equation}
\kappa_{\gamma\,mod}=\frac{n_{X\,mod}-n_{\bar{X}\,mod}}{n_{\gamma\,mod}}.
\label{kappagamma}
\end{equation}
However for practical purposes it is more suitable to use the ratio to entropy density which,
unlike Eq.(\ref{kappagamma}), does not change in time provided entropy conservation.
Photon number density $n_{\gamma}$ and entropy density $s$ are given by
\begin{equation}
n_{\gamma}=\frac{2\zeta(3)}{\pi^2}T^3,\;\;\; s=\frac{2\pi^2 g_s}{45}T^3=1.80g_sn_{\gamma},
\label{ngammas}
\end{equation}
where
\begin{equation}
g_s=\sum_{bos} g_S(\frac{T_{bos}}{T})^3+\frac{7}{8}\sum_{ferm}g_S(\frac{T_{ferm}}{T})^3.
\label{gs}
\end{equation}
The sums in Eq.(\ref{gs}) are over ultrarelativistic bosons and fermions.
So
\begin{equation}
\kappa=\frac{n_{X}-n_{\bar{X}}}{s},\;\;\;
\kappa=\frac{\kappa_{\gamma\,mod}}{1.8g_{s\,mod}},
\label{kappa}
\end{equation}
$g_{s\,mod}\approx 3.93$.

Eq.(\ref{kappa}) provides connection between $n_{X}$ and $n_{\bar{X}}$.
Let us pass to the variables
\begin{equation}
r_+=\frac{n_X}{s},\;\;\; r_-=\frac{n_{\bar{X}}}{s},\;\;\; r=\frac{n_X+n_{\bar{X}}}{s},\;\;\;
x=\frac{T}{m}.
\label{rx}
\end{equation}
Apparent relations between $r_i$ are
\begin{equation}
r_+-r_-=\kappa,\;\;\; r_++r_-=r.
\label{r-r-r}
\end{equation}

Provided that essential entropy re-distribution does not take place ($g_s=const$) during the period of freezing out, transformation to variable $x$ is possible $$-Hdt=dT/T=dx/x.$$
On the RD stage Hubble constant depends on $T$ as
\begin{equation}
H=\frac{2\pi}{3} \sqrt{\frac{\pi g_{\epsilon}}{5}} \frac{T^2}{m_{Pl}},
\label{Heps}
\end{equation}
where $g_{\epsilon}$ is given by
\begin{equation}
g_{\epsilon}=\sum_{bos} g_S(\frac{T_{bos}}{T})^4+\frac{7}{8}\sum_{ferm}g_S(\frac{T_{ferm}}{T})^4.
\label{geps}
\end{equation}
For $r_+$, $r_-$ and $r$ from Eqs.(\ref{asym}) one obtains equations
\begin{eqnarray}
\frac{dr_+}{dx}=f_1\left<\sigma_{ann}v\right>\left( r_+(r_+-\kappa)-f_2(x)\right)\nonumber\\
\frac{dr_-}{dx}=f_1\left<\sigma_{ann}v\right>\left( r_-(r_-+\kappa)-f_2(x)\right)\nonumber\\
\frac{dr}{dx}=\frac{1}{2}f_1\left<\sigma_{ann}v\right>\left( r^2-\kappa^2-4f_2(x)\right).
\label{drrr}
\end{eqnarray}
Here
\begin{eqnarray}
f_1=\frac{s}{Hx}\nonumber\\
f_2(x)=\frac{n_{eq}^2}{s^2}=\frac{45^2 g_S^2}{2^5\pi^7g_s^2 x^3}\exp\left(-\frac{2}{x}\right).
\label{f12}
\end{eqnarray}
With the use of Eqs.(\ref{ngammas}) and Eq.(\ref{Heps}) one finds that on the RD stage $f_1$ is equal to
$$f_1=\sqrt{\frac{\pi g_s^2}{45g_{\epsilon}}}m_{Pl}m$$
and independent of $x$.

To solve Eqs.(\ref{drrr}) analogously to Eq.(\ref{sym}), namely neglecting $f_2(x)$ in them starting with some $x=x_f$,
it would not be more difficult if to define the moment $x=x_f$.

Nonetheless, if one supposes that such a moment is defined then, say, $r_i$ will be
\begin{eqnarray}
r_+(x\approx 0)=\frac{\kappa\cdot r_{+f}}{r_{+f}-(r_{+f}-\kappa) \exp\left(-\kappa J\right)}\nonumber\\
r_-(x\approx 0)=\frac{\kappa\cdot r_{-f}}{(\kappa+r_{-f}) \exp\left( \kappa J \right)-r_{-f}}\\
r(x\approx 0)=\kappa \frac{(\kappa+r_{f})\exp\left( \kappa J \right)+r_f-\kappa}
{(\kappa+r_{f})\exp\left( \kappa J \right)-(r_f-\kappa)}.\nonumber
\label{rrr}
\end{eqnarray}
Here $r_{i\,f}=r_i(x=x_f)$,
$$J=\int_0^{x_f} f_1\left<\sigma_{ann}v\right>dx.$$
All $r_i$ (at any moment) are related with the help of Eqs.(\ref{r-r-r}). Taking into account Eq.(\ref{nmueq}) or Eq.(\ref{neq2})
for $r_{i\,f}$ one obtains
\begin{eqnarray}
r_{\pm\,f}=\frac{1}{2}\left(\sqrt{4f_2(x_f)+\kappa^2}\pm \kappa \right),\;\;\;
r_{f}=\sqrt{4f_2(x_f)+\kappa^2}.
\label{rpm}
\end{eqnarray}
For $\left<\sigma_{ann}v\right>$ independent of $x$ on RD stage, when $f_1$ is also independent of $x$, with the account for the definition of $x_f$ from the condition  $R(T_f)=H(T_f)$ for reaction rate $R(T_f)=n_{eq}(T_f)\left<\sigma_{ann}v(T_f)\right>$, leading to
$$n_{eq}(T_f)\left<\sigma_{ann}v(T_f)\right>/H(T_f)=\frac{n_{eq}}{s}\cdot \frac{s}{H x_f}\cdot \left<\sigma_{ann}v(x_f)\right> \cdot x_f =$$
\beq
= \sqrt{f_2(x_f)} f_1 \left<\sigma_{ann}v(x_f)\right> \cdot x_f =1,
\label{f2f1H}
\eeq
one obtains
\beq
\sqrt{f_2(x_f)} = \frac{1}{f_1 \left<\sigma_{ann}v\right> \cdot x_f} = \frac{1}{J}.
\label{f2f1J}
\eeq

If (a) $\left<\sigma_{ann}v\right>=\alpha^2/m^2$ or (b) $\left<\sigma_{ann}v\right>=C\alpha/\sqrt{Tm^3}$
and one assumes $f_1=const$ then
\begin{eqnarray}
J_a=\sqrt{\frac{\pi g_s^2}{45g_{\epsilon}}}m_{Pl}\frac{\alpha^2}{m}x_f, \nonumber\\
J_b=\sqrt{\frac{\pi g_s^2}{45g_{\epsilon}}}m_{Pl}C\frac{\alpha}{m}2\sqrt{x_f}.
\end{eqnarray}

In the case of freezing out of $U$quarks one has
$$f_{1U}=\sqrt{\frac{\pi g_s^2}{45g_{\epsilon}}}m_{Pl}m_U \approx 2.5 m_{Pl}m_U,$$
$\left<\sigma_{ann}v\right> =\frac{1}{N_c}\frac{\bar \alpha^2}{m^2}$ and
\beq
J_U=\sqrt{\frac{\pi g_s^2}{45g_{\epsilon}}}m_{Pl}\frac{1}{N_c}\frac{\bar \alpha^2}{m}x_f,
\label{JU}
\eeq
where $\bar \alpha=C_F\alpha_s  \sim (4/3) \cdot 0.1 \approx 0.13$
and $C_F=(N_c^2-1)/2N_c=4/3$ is the color factor.
Another color factor $1/N_c=1/3$
is the probability to find an appropriate anticolour. Putting in Eq.(\ref{f12}) $g_S=6$, $g_s \sim 100$, one obtains solution of transcendent equation (\ref{f2f1J})
$$x_f \approx \left(\ln{\left(\frac{45 g_S}{2^{5/2}\pi^{7/2}g_s} \cdot f_{1U} \left<\sigma_{ann}v\right>\right)}\right)^{-1} \approx$$
$$ \approx \frac{1}{30}\cdot \frac{1}{(1 - \ln{(S_6)}/30)}.$$
Taking $g_s \approx g_{\epsilon} \sim 100$
one finds from Eq.(\ref{JU}) $J_U = 1.3 \cdot 10^{12}/S_6 (1 - \ln{(S_6)}/30)^{-1}$ and from Eq.(\ref{f2f1J}) $\sqrt{4f_2(x_f)} = 2/J_U =16 \cdot 10^{-13} S_6\cdot (1 - \ln{(S_6)}/30)$. For $\kappa = r_U= 1.2 \cdot 10^{-13}/S_6$
one has $\kappa J_U = 0.16/S_6^2$.
one obtains. Since $4f_2(x_f) \gg \kappa^2$ for $S_6 \ge1$ one obtains from Eq.(\ref{rpm})
\beq
r_{\pm\,f}=\frac{1}{2}\left(\sqrt{4f_2(x_f)}\pm \kappa \right).
\label{rpmf}
\eeq
It gives for the frozen out abundances of $U$ and $\bar U$
\begin{eqnarray}
r_U=\frac{\kappa\cdot r_{+f}}{r_{+f}-(r_{+f}-\kappa) \exp\left(-\kappa J_U\right)}\nonumber\\
r_{\bar U}=\frac{\kappa\cdot r_{-f}}{(\kappa+r_{-f}) \exp\left( \kappa J_U \right)-r_{-f}}.
\label{rUpm}
\end{eqnarray}
With the account for the numerical values taken above one gets $r_U \approx 8.6 \cdot 10^{-13}$
and $r_{\bar U} \approx 7.4 \cdot 10^{-13}$ for $S_6=1$. For growing $S_6 > 1$ the solution Eq.(\ref{rUpm}) approaches the values
\begin{eqnarray}
r_U \approx \sqrt{f_2(x_f)} + \kappa/2 \approx\nonumber\\
\approx 8 \cdot 10^{-13}S_6\cdot (1 - \ln{(S_6)}/30) + 6 \cdot 10^{-14}/S_6\nonumber\\
r_{\bar U} \approx \sqrt{f_2(x_f)} - \kappa/2 \approx \nonumber\\
\approx 8 \cdot 10^{-13}S_6\cdot (1 - \ln{(S_6)}/30) - 6 \cdot 10^{-14}/S_6.
\label{rUSpm}
\end{eqnarray}
At $S_6 < 0.4$ the factor in exponent $\kappa J_U$ exceeds 1, and some suppression of $\bar U$ abundance takes place. For $S_6=0.2$ it reaches maximal possible value $\kappa J_U=4$ and the solution Eq.(\ref{rUpm}) gives $r_U \approx \kappa = 6 \cdot 10^{-13}$, $r_{-f} \approx 1.1 \cdot 10^{-14}$ from Eq.(\ref{rpm}) and
$$r_{\bar U} \approx \frac{\kappa\cdot r_{-f}}{\kappa+r_{-f}}\exp\left(- \kappa J_U \right) \approx 1.1 \cdot 10^{-14} \exp\left( -4 \right) \approx 2 \cdot 10^{-16}.$$

In the case of freezing out of $E$-leptons one has $f_{1E} \approx 2.5 m_{Pl}m_E,$
$\left<\sigma_{ann}v\right> \approx \frac{\alpha^2}{m_E^2}$ and
\beq
J_E=\sqrt{\frac{\pi g_s^2}{45g_{\epsilon}}}m_{Pl}\frac{\alpha^2}{m_E}x_f.
\label{JE}
\eeq
Putting in Eq.(\ref{f12}) $g_S=2$, $g_s \sim 100$, one obtains solution of transcendent equation (\ref{f2f1J})
$$x_f \approx \left(\ln{\left(\frac{45 g_S}{2^{5/2}\pi^{7/2}g_s} \cdot f_{1E}\left<\sigma_{ann}v\right>\right)}\right)^{-1} \approx$$
$$ \approx \frac{1}{25}\cdot \frac{1}{(1 - \ln{(S_6)}/25)}.$$
Taking $g_s \approx g_{\epsilon} \sim 100$, one finds from
Eq.(\ref{JE}) $J_E \approx (10^{11}/S_6) \cdot (1 -
\ln{(S_6)}/25)^{-1}$. For $\kappa = r_E= 8 \cdot 10^{-14}/S_6$ it
corresponds to $\kappa J_E \approx 8 \cdot 10^{-3} /S_6^2\ll 1$
for all $S_6 > 0.2$ and one obtains $\sqrt{4f_2(x_f)} = 2/J_E= 2
\cdot 10^{-11} S_6\cdot (1 - \ln{(S_6)}/25)$. Since $4f_2(x_f) \gg
\kappa^2$ one obtains from Eq.(\ref{rpm}) \beq
r_{\pm\,f}=\frac{1}{2}\left(\sqrt{4f_2(x_f)}\pm \kappa \right).
\label{rpmf1} \eeq For small  $\kappa J_E \ll 1$ frozen out
abundances of $E$ and $E^+$ have the form
\begin{eqnarray}
r_E \approx \frac{r_{+f}}{1+ (r_{+f}-\kappa) J_E} \approx \sqrt{f_2(x_f)} + \kappa/2\nonumber\\
r_{E^+} \approx \frac{r_{-f}}{1+ (\kappa+r_{-f}) J_E} \approx \sqrt{f_2(x_f)} - \kappa/2.
\label{rEpm}
\end{eqnarray}
For the numerical values taken above and $S_6=1$ one gets $r_E = 0.996 \cdot 10^{-11}$
and $r_{E^+}= 1.004 \cdot 10^{-11}$. In case of minimal possible value $S_6=0.2$ $r_E = 2.2 \cdot 10^{-12}$ and $r_{E^+}= 1.8 \cdot 10^{-12}$.

\section*{\label{radiative} Appendix 2. Recombination and binding of heavy charged particles.}

In the analysis of various recombination processes
we can use the interpolation formula for recombination cross section,
deduced in  \cite{4had}:
\beq
 \sigma_r=(\frac{2\pi}{3^{5/2}}) \cdot \frac{\bar \alpha ^3}{T\cdot I_1} \cdot \log{(\frac{I_1}{T})}\eeq
and the recombination rate given by \beq
 \sv=(\frac{2\pi}{3^{5/2}}) \cdot \frac{\bar \alpha ^3}{T\cdot I_1} \cdot \log{(\frac{I_1}{T})} \cdot \frac{k_{in}}{M}
\label{recdisc} \eeq Here $k_{in}= \sqrt{2 T M}$, $I_1 \approx
\bar \alpha^2 M/2$ is the ionization potential and $M$ has the
meaning of the reduced mass for pair of recombining particles.
Pending on the process, the constant $\bar \alpha$ has the meaning
of fine structure constant $\alpha$ or QCD constant $\alpha_c$.
The approximation Eq.(\ref{recdisc}) followed from the known
result for
 the electron-proton recombination
\beq \sigma_{rec}=\sigma_r
  =\sum_i \frac{1}{N_c} \frac{8\pi}{3^{3/2}} \bar \alpha^3 \frac{e^4}{Mv^2i^3} \frac{1}{(Mv^2/2+I_i)},
\label{recep}
\eeq
  where
 $M$ and $v$ are the reduced mass and velocity of particles;
 $I_i$ - ionization potential  ($I_i=I_1/i^2$).  The color factor $1/N_c=1/3$
is the probability to find an appropriate anticolor.

 To sum approximately over 'i' it was noted in  \cite{4had} that $\sigma_r\propto 1/i$
 for $I_i >> Mv^2/2=T_{eff}$ while at $I_i<T_{eff}$ the cross section
 $\sigma_i\propto 1/i^3$ falls down rapidly.

The following classical description is valid for $v/c \ll \alpha$.

Radiation of mutually attracting opposite charges
can lead to formation of their bound system. It
can be described \cite{4had} in the analogy to the process of free
monopole-antimonopole annihilation considered in \cite{ZK}.

Potential energy of Coulomb interaction between opposite charges
exceeds their thermal energy $T$ at the distance
$$ d_0 = \frac{\alpha}{T}.$$
According to \cite{ZK}, following the classical
solution of energy loss due to radiation, converting infinite
motion to finite, free charged particles form bound
systems at the impact parameter \beq a \approx (T/m)^{3/10} \cdot
d_0. \label{impact} \eeq The rate of such binding is then given by
\beq \sv = \pi a^2 v \approx \pi \cdot (m/T)^{9/10} \cdot
(\frac{\alpha}{m})^2 .\eeq

The successive evolution of this highly excited atom-like bound
system is determined by the loss of angular momentum owing to
radiation. The time scale for the fall on the center in this
bound system can be estimated
according to classical formula (see \cite{DFK})

\beq \tau = \frac{a^3}{64 \pi} \cdot (\frac{m}{\alpha})^2 =
\frac{\alpha}{64 \pi} \cdot (\frac{m}{T})^{21/10} \cdot
\frac{1}{m} .\label{recomb} \eeq

As it is easily seen from Eq.(\ref{recomb}) this  recombination time scale
$\tau \ll m/T^2 \ll m_{Pl}/T^2$ turns to be
much less than the cosmological time at which the bound system was
formed.

Classical description assumes $a= \frac{\alpha}{m^{3/10}T^{7/10}} \gg  \frac{1}{\alpha m}$ and is valid at $T \ll m  \alpha^{20/7}$.

\section*{\label{recQ} Appendix 3. Elimination of $\bar U$ in $\bar U U$bound systems}
When temperature falls down below $I_U \approx \bar \alpha_c^2 M_U/2 \sim 15 GeV S_6$
(where $\bar \alpha= 4/3 \cdot 0.1 \approx 0.13$ and $M_U = m_U/2$ is the reduced mass of $U$ quarks in $UU$ system - see Appendices 1 and 2)
$U$-quarks begin to bind due to chromo-Coulomb attraction.
They form bound $(UU)$ diquark systems and $(UUU)$ baryons. Similar to $^4$He formation in SBBN, $(UUU)$, being the system with the largest binding energy, is not produced by 3-body process directly, but by multi-step 2-body events.  In SBBN the chain of nucleosynthesis reactions starts with formation of D, and all the frozen out neutrons are first bound in D. In analogy the process of $U$ binding starts with formation of $(UU)$ diquarks.

Simultaneously at $T< I_U \approx \bar \alpha_c^2 m_U/4 \sim 15 GeV S_6$ the frozen out antiquarks $\bar U $ begin to bind with $U$
quarks into charmonium-like state $(\bar U U)$
and annihilate. However there is an important difference between formation of $(UU)$ diquarks and $(\bar U U)$ charmonium-like systems. The former are stable relative to annihilation and can be disrupted by energetic gluons, while the latter annihilate on the timescale (see Appendix 2), much less than the timescale of gluon interaction. Therefore direct reaction of $(\bar U U)$ is not compensated by inverse process of its disruption by energetic gluons, and formation of $U \bar U$ systems and $\bar U$ annihilation in them starts immediately after temperature falls down below $I_U$.

The decrease of $\bar U$ abundance owing to $U \bar U$
recombination is governed by the equation
\beq
\frac{dr_{\bar U}}{dt} = -r_U r_{\bar U} \cdot s \cdot \sv,
\label{hadrecomb}
\eeq
where $s$ is the entropy density and (see \cite{4had})
\beq
\sv \approx  (\frac{16\pi}{3^{5/2}}) \cdot \frac{\bar \alpha}{T^{1/2}\cdot m_U^{3/2}}.
\label{svUbar}
\eeq

With the use of formulae in Appendix 1
Eq.(\ref{hadrecomb}) is reduced to the form:
\beq
\frac{dr_{\bar U}}{dx}=f_1\left<\sigma v\right> r_{\bar U}(r_{\bar U}+\kappa_U),
\label{Usrecomb}
\eeq
where $x=T/I_U$, the asymmetry $\kappa_U =  r_U-r_{\bar U} =1.2 \cdot 10^{-13} /S_6$ is given by Eq.(\ref{Uexcess}) and
$$f_{1U}=\sqrt{\frac{\pi g_s^2}{45g_{\epsilon}}}m_{Pl}I_U \approx 2.5 m_{Pl}I_U.$$
The concentration of remaining $\bar U$ is given by
\beq
r_{\bar U}=\frac{\kappa_U\cdot r_{\bar U f}}{(\kappa_U+r_{\bar U f})} \exp\left( -\kappa J \right),
\label{rUbar}
\eeq
where from Eq.(\ref{rUfr}) $r_{\bar U f} = 7.4 \cdot 10^{-13} f_{\bar U} (S_6)$, and
\beq
J=\int_0^{x_f} f_{1U}\left<\sigma v\right>dx = 4 \cdot 10^{14}/S_6.
\label{JUbar}
\eeq
It was taken into account in Eq.(\ref{JUbar}) that in the considered case annihilation starts at $T\sim I_U$ so that $x_f \sim 1$.
One obtains that $\bar U$ are practically eliminated at $T \sim I_U \approx \bar \alpha_c^2 M_U/2 \sim 15 GeV S_6$, since their abundance decreases down to
$r_{\bar U} \approx 1 \cdot 10^{-13} \cdot \exp\left( -48/S_6^2 \right),$ being $\sim 7 \cdot 10^{-34}$ for $S_6=1$.
Therefore at $S_6\le 1$ their role is negligible in the successive processes. That is not the case for $E^+$ for the same small values of $S_6$.
In the latter case, which is described by similar equation with similar solution, recombination rate, having the form Eq.(\ref{svUbar}), involves  fine structure constant $\alpha$ instead of QCD constant $\bar \alpha$. It makes the corresponding value of $J$ (with the account for other numerical factors: difference in masses of $U$ and $E$ and in statistical factors in the period of their binding, color factors) about 90 times smaller and the exponential suppression is practically absent for $E^+$ at $S_6 \ge 1$ (see subsection \ref{anE}).  This makes the $E^+$ annihilation an aborted one at $S_6 \ge 1$ and it leaves tera-positron as a remarkable imprint of the sinister Universe in later days.

Note that at $S_6 > 7$ exponential suppression is virtually absent for $\bar U$ too.
\section*{\label{recU} Appendix 4. $U$ recombination into $(UU)$ and $(UUU)$ systems}
At $T< I_U \approx \bar \alpha_c^2 M_U/2 \sim 15 GeV S_6$ free $U$ can bind into $(UU$) diquarks, but during long period direct reaction
\beq
U+U \rightarrow (UU) +g,
\label{dir1}
\eeq
where $g$ is gluon is balanced by the inverse process
\beq
g +(UU)\rightarrow U+U.
\label{inv1}
\eeq
It reminds the well known "entropy barrier" for $n+p \rightarrow D + \gamma$ reaction in SBBN.
Since relative abundance of gluons $r_g \sim 0.1 \gg r_U$, gluons can effectively distract $(UU)$ even at temperatures $T \ll I_U$. It provides the conditions for kinetic equilibrium between direct and inverse reactions.

As soon as $(UU)$ are formed the processes
\beq
U+(UU) \rightarrow (UUU) +g,
\label{dir2}
\eeq
and
\beq
U+U \rightarrow (UU) +g,
\label{dir3}
\eeq
are possible.

In equilibrium abundance of these bound systems is
determined by Saha equations
\beq
\frac{n_U n_U}{n_g n_{(UU)}} = \exp{(-\frac{I_{U2}}{T})}.
\label{recSaha1}
\eeq
and
\beq
\frac{n_U n_{(UU)}}{n_g n_{(UUU)}} = \exp{(-\frac{I_{U3}}{T})}.
\label{recSaha2}
\eeq
At $T< T_{0U} \approx 1/30 I_U \approx 0.5 GeV S_6$  (corresponding to period $t_{0U} \sim 4 \cdot 10^{-6}$s$/\sqrt{S_{6}}$) fraction of free $U$quarks and $(UU)$diquarks
begins to decrease being governed by the system of kinetic equations.
Solution of these equations for free $U$ and $(UU)$ requires development of special system of equations and their proper numerical treatment. However the precise result for the considered set of reactions is not so important, since it can be strongly modified and washed out by the successive processes of hadronic recombination considered in Appendix 5.

Qualitatively one may conclude that the most of initially free $U$ bind into $(UUU)$ systems, which contain the bulk of the $B'$ excess, so that $r_{UUU} = r_{B'} = 4 \cdot 10^{-14}/S_6$. However, the residual amount of unbound $U$ and $(UU)$ can not be small, since their binding into $(UUU)$ stops, when $n_U \sv t_{0U} \sim 1$ and $n_{UU} \sv t_{0U} \sim 1$. These conditions are realized at $S_6 \sim 1$ for
\beq
\frac{n_U}{n_{(UUU)}} \sim \frac{n_{(UU)}}{n_{(UUU)}} \sim \frac{1}{10}.
\label{recfrU}
\eeq

\section*{\label{recQbar} Appendix 5. Formation of $(UUU)$ due to hadronic recombination}
After QCD phase transition at $T = T_{QCD} \approx 150$MeV free $U$ quarks
and $UU$ diquarks combine with light quarks into $(Uqq)$ and $(UUq)$ hadrons. In
baryon asymmetrical Universe only excessive valence quarks should
enter such hadrons, so that $U$-baryons are formed, while
the abundance of $(U \bar q)$ mesons
is suppressed exponentially. These details of $U$-quark hadronization are
discussed in  \cite{4had}.

Since $(Uqq)$ and $(UUq)$ baryons have hadronic size, their collisions with typical hadronic cross sections can provide additional hadronic recombination of $U$ and $UU$ into $(UUU)$  \cite{Glashow}. However the analysis of this problem has revealed a substantial uncertainty in the estimation of recombination rate \cite{4had}.

 The maximal estimation for the
reaction rate of recombination $\sv$ is given by
\beq
\sv \sim \frac{1}{m_{\pi}^2} \approx 6 \cdot 10^{-16}\, {\rm
\frac{cm^3}{s}} \label{hadsigmv}
\eeq
or by
\beq
\sv \sim
\frac{1}{m_{\rho}^2} \approx 2 \cdot 10^{-17} \frac{\rm cm^3}{\rm s}.
\label{hadsigmv1}
\eeq
These estimations assume that in the process of collision recombination takes place due to emission of light quarks, what has typical hadronic cross section.

The minimal estimation (see \cite{4had}) was based on the QCD consideration, assuming that the process of $U$ binding takes place at small distances and is not influenced by effects of QCD confinement. It gives recombination rate  \cite{4had}
\beq \sv \approx 0.4 \cdot (T_{eff}m^3)^{-1/2}
(3 + \ln{(T_{QCD}/T_{eff})}).
\label{hadrecmin}
\eeq
Here $T_{eff}= \max{\{T, T_b\}}$ takes into account that at $T<T_b\approx \frac{m_{light}^2}{2m} \approx 1.5 \cdot  10^{-2}\MeV\frac{3.5\TeV}{m}$,
where $m_{light}\approx 300$ MeV is the constituent mass of the light quark of hadron, kinetic energy
of recombining quarks is determined by their motion inside hadrons.

Binding of $U$ in the course of hadronic recombination takes place in reactions \cite{Glashow}
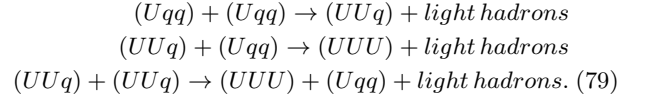
\begin{eqnarray}
(Uqq) + (Uqq) \rightarrow (UUq) + light\, hadrons\nonumber\\
(UUq) + (Uqq) \rightarrow (UUU) + light\, hadrons\nonumber\\
(UUq) + (UUq) \rightarrow (UUU) + (Uqq) + light \,hadrons.
\label{hadrUbind}
\end{eqnarray}
Detailed analysis of these processes needs special numerical treatment. However such detailed description can hardly change the qualitative result that the remaining amount of $(Uuu)$ and $(UUu)$ can not be too small.

Assuming $r_{(UUu)} = r_{(Uuu)} = r_h$ one can describe reactions Eq.(\ref{hadrUbind}) by equation
\beq
\frac{dr_{h}}{dx}=f_{1h}\left<\sigma v\right> r_{h}^2,
\label{Ushadrec}
\eeq
where $x=T/T_{QCD},$
$$f_{1h}=\sqrt{\frac{\pi g_s^2}{45g_{\epsilon}}}m_{Pl}T_{QCD} \approx m_{Pl}T_{QCD}.$$
and $\sv$ is given by Eqs.(\ref{hadsigmv}), (\ref{hadsigmv1}) or
(\ref{hadrecmin}). Solution of the equation Eq.(\ref{Ushadrec}) is given by
\beq
r_{h}=\frac{r_{h f}}{1+r_{h f}J_h},
\label{rUhadrec}
\eeq
where from Eq.(\ref{recfrU})
$r_{h f} = 0.1 r_{B'} = 4 \cdot 10^{-15}/S_6$ and
\beq
J_h=\int_0^{x_f} f_{1h}\left<\sigma v\right>dx.
\label{JUhadrec}
\eeq
Since reactions Eq.(\ref{hadrUbind}) start immediately after QCD phase transition at $T \sim T_{QCD}$, in Eq.(\ref{JUhadrec}) $x_f \sim 1$.
Depending on the choice of  $\sv$ the remaining amount of $(Uuu)$ and $(UUu)$ ranges from the case A to case C.
Case A.

The value of $\sv$ is equal to Eq.(\ref{hadsigmv}). Then
$$J_{hA} =  m_{Pl}T_{QCD} \frac{1}{m_{\pi}^2} \approx 10^{20}.$$
In the solution Eq.(\ref{rUhadrec}) $r_{h f}J_{h A} = 4 \cdot 10^{5}/S_6 \gg 1$ at $S_6 \ll 4 \cdot 10^5$ and the remaining amount of $(Uuu)$ and $(UUu)$ is independent on their initial abundance, being equal to
\beq
r_{h A} = \frac{1}{J_{h A}} \approx 1.0\cdot 10^{-20}.
\label{hadrecsol}
\eeq

Case B

For $\sv$ from Eq.(\ref{hadsigmv1}) $J_{h B}$ is $(\frac{m_{\rho}}{m_{\pi}})^2 \sim 30$ times smaller than $J_{h A}$. It results in correspondingly $30$ times larger amount of $(Uuu)$ and $(UUu)$, which is now valid for $S_6 \ll 10^4$:
\beq
r_{h B} = \frac{1}{J_{h B}} \approx 3.0 \cdot 10^{-19}.
\label{hadrecsol2}
\eeq

Case C

The minimal estimation of recombination rate Eq.(\ref{hadrecmin}) leads to relatively small value of $J_h$:
$$J_{h C} = 2 \frac{m_{Pl}}{m_{U}} (\frac{T_{QCD}}{m_{U}})^{1/2} \approx 4 \cdot 10^{13}/S_6^{3/2}$$
and the solution of Eq.(\ref{rUhadrec}) practically coincides with the initial value $r_{h f}$
\beq
r_{h C} = \frac{r_{h f}}{1 + r_{h f}J_{h C}} \approx r_{h f}
\approx 4\cdot 10^{-15}/S_6.
\label{hadrecsol3}
\eeq
However, even in the case C the abundance of $(Uuu)$ and $(UUu)$ is smaller at $S_6 \ge 1$, than of $(UUU)$. The product $r_{h f}J_{h C}$ grows at small $S_6 < 1$ as $\propto S_6^{-5/2}$
and reaches the value $\sim 9$ at minimal allowed value $S_6=0.2.$ That leads to corresponding order of magnitude decrease of $r_h$ in the case C.

On the other hand, the residual amount of $(Uuu)$ and $(UUu)$ in the most optimistic case A  is no less than $2.5\cdot 10^{-7}$ of $(UUU)$ relative to baryons.
This fact reveals a potential danger for the sinister model. Even being bound with $EE$ these hadrons are not elusive: their interaction with matter has normal hadronic cross section.

Note that at $S_6 > 7$ hadronic recombination suppresses abundance of primordial "tera-mesons" $(\bar U u)$, which were not suppressed in $U$-quark recombination. In case A suppression is $\propto \exp{\left(-r_{h f}J_{h A}\right)}=\exp{\left(-\frac{4 \cdot 10^{5}}{S_6}\right)}$ and in case B
$\propto \exp{\left(-\frac{1.3 \cdot 10^{4}}{S_6}\right)}$. In the case C, on the other hand, $r_{h f}J_{h C}=0.16/S_6^{-5/2}$ and there is no additional suppression for $S_6 \ge 1.$
%
%
\section*{\label{rece} Appendix 6. Complete recombination of charged tera particles}
Cosmological abundance of free charged $U$-baryons is to be exponentially
small after recombination. If the lightest is $(Uuu)$ baryon with
electric charge $+2$, atoms of anomalous He are formed
by it as well as by free $(UUu)$ and $(UUU)$ baryons.
Their recombination takes place together with ordinary He recombination
at $T < I_{He} = 54.4 eV$.  Taking the equation for the residual amount
of free ions in the form
\beq
\frac{dr_{i}}{dx} =f_{1He} r_{i}r_{e} \frac{\sv_{2rec}^0}{x^{1/2}},
\label{recHel}
\eeq
where $x=T/I_{He}$, $f_{1He} \approx m_{Pl}I_{He}$, $r_e=r_p \approx r_b= 0.8 \cdot 10^{-11}$,
$$\sv_{2rec}^0 = \sv_{2rec}x^{1/2}=(\frac{4\pi}{3^{3/2}}) \cdot \frac{Z^2 \alpha^2}{I_{He}\cdot m_e},$$
charge of He $Z=2$, we find that the solution \beq r_{i}=r_{i0}
\exp\left( -r_{e} J_{He} \right) \label{recHesol} \eeq with \beq
J_{He}=\int_0^{x_{He0}} f_{1Ep}\left<\sigma v\right>_{2rec}dx =
\frac{m_{Pl}}{m_e}(\frac{8\pi}{3^{3/2}}) Z^2 \alpha^2 \cdot
\sqrt{x_{He0}} \eeq where $x_{He0} \sim 1/30$, contains huge
negative number ($-r_{e} J_{He} \approx -3 \cdot 10^8$) in
exponent. $(^4HeE^-)^+$, $E^+$, $U$-hadrons and systems with
charge $+1$ form atoms of anomalous hydrogen. Their recombination
with $e$ takes place together with ordinary hydrogen on MD stage
at $T \sim I_H/30<T_{RM} \approx 1 eV$, where $I_H =13.6 eV$. The
form of equation for decrease of free ion abundance is similar to
Eq.(\ref{recHel}) and reads \beq \frac{dr_{i}}{dx} =f_{1H}
r_{i}r_{e} \frac{\sv_{1rec}^0}{x^{1/2}}, \label{recHel1} \eeq
where $x=T/I_{H}$, $f_{1H} \approx m_{Pl}I_{H}\frac{x}{x_{RM}}$,
$$\sv_{1rec}^0 = \sv_{1rec}x^{1/2}=(\frac{4\pi}{3^{3/2}}) \cdot \frac{ \alpha^2}{I_{H}\cdot m_e}.$$
The solution
\beq
r_{i}=r_{i0} \exp\left( -r_{e} J_{H} \right)
\label{recHsol}
\eeq
has the form of Eq.(\ref{recHesol}) with
\beq
J_{He}=\int_0^{x_{H0}} f_{1Ep}\left<\sigma v\right>_{2rec}dx
=  \frac{m_{Pl}}{m_e}(\frac{2\pi}{3^{3/2}})  \alpha^2 \cdot x_{H0}/\sqrt{x_{RM}}
\eeq
and also contains huge negative number ($-r_{e} J_{H} \approx -10^7$) in exponent.

\section*{\label{stars} Appendix 7. Gravitational concentration inside stars}
For number density $n_s$ of stars with mass $M_s$ and radius $R_s$ the decrease of  number density $n_i$ of free particles, moving with relative velocity $v$, is given by
\beq
\frac{dn_i}{dt} = -n_s n_i \pi R_s(R_s + \frac{2 G M_s}{v^2})v \approx -n_s n_i 2\pi \frac{2 G M_s}{v}.
\eeq
Therefore, to be effective (i.e. to achieve substantial decrease of number density $n_i=n_{i0} \exp(-t/\tau)$)
the timescale of capture
$$\tau = \frac{1}{n_s 2 \pi R_s G M_s/v}$$
should be much less than the age of the Universe $\tau \ll t_U= 4\cdot 10^{17}$s, whereas for $n_s \sim 1 pc^{-3}$, $M_s = M_{\odot} \approx 2 \cdot 10^{33}$g, $R_s = R_{\odot} \approx 7 \cdot 10^{10}$cm and $v\sim10^{6}cm/s$, $\tau \sim 5 \cdot 10^{23} s \gg t_U$. Even for supergiants with $M_s \sim 20 M_{\odot}$
and $R_s \sim 10^4 R_{\odot}$ (and even without account for smaller number density of these stars) we still obtain $\tau \sim 3 \cdot 10^{18} s \gg t_U$.

\section*{\label{accelerators} Appendix 8. Signatures for tera-particles in Lab }
 In the discussion of this problem we use the results of \cite{4had}.
The assumed values of $E$ and $U$-quark masses make the problem of their search at accelerators similar
to the case of other heavy quark. However, the strategy of such search should take into account
the principal difference from the case of unstable  quark (e.g. top-quark). One should expect that
in the considered case a stable particle should be produced.
Note that $UUU$ $\bar U \bar U\bar U$ pair production is beyond the reach of the
next generation of colliders, as well as the probability for production of such
pair is strongly suppressed. So it is the pair of $EE^+$ or single $U$ and $\bar U$ - hadrons,
what can be expected in colliders above their threshold.

In the case of $E E^+$ pair two charged stable leptons are produced.
In the case of $U \bar U$ dominantly a pair of mesons $U \bar q$ and $\bar U q$ appears.
The relative probability is $<0.1$ for creation of a baryon pair $(Uqq)$ and $(\bar U \bar q \bar q)$ (see \cite{4had}).

Charged heavy stable particles can be observed as the 'disagreement' between the
track curvature (3-momentum)
\beq
p = 0.3 B \cdot R \cdot Q,
\label{curv}
\eeq
and the energy of the track measured in the calorimeter (or energy
loss $dE/dx$). In Eq.(\ref{curv}) $B$ is magnetic field in $T$, $p$ is
momentum in GeV, $R$ is radius of curvature in meters, $Q$ is the
charge of particle in the units of elementary charge $e$,


Due to a very large mass the created heavy tera-particles are
rather slow. About half of the yield is given by particles with the
velocity $\beta < 0.7$. To identify such
particles one may study the events with a large transverse energy (say,
using the trigger - $E_T > 30$ GeV). The signature for a new heavy
hadrons will be the 'disagreement' between the values of the full
energy $E=\sqrt{m^2+|\vec p|^2}-m$ measured in the calorimeter, the
curvature of the track (which, due to a larger momentum $|\vec
p|=E/\beta$, will be smaller than that for the light hadron where
$E\simeq |p|$) and the energy loss $dE/dx$.
For the case of heavy hadrons due to a low $\beta$
 the energy loss $\frac{dE^{QED}}{dx}$ caused by the electromagnetic interaction
 is larger than that for the ultrarelativistic light hadron, while
 in "hadron calorimeter" the energy loss caused by the strong
 interactions is smaller (than for a usual light hadron), due to a
 lower inelastic cross section for a smaller size heavy hadron,
 like $(U\bar d)$ meson.
Besides this the whole large $E_T$ will be produced by the single
isolated track and not by a usual hadronic jet, since the expected
energy of the accompanying light hadrons
 $E_T^{acc}\sim \frac{1\  GeV}{m}E_T$ is rather low.

Another possibility to identify the new stable heavy hadrons is
 to use the Cherenkov counter or the time-of-flight information.

We hope that tera-particles, if they exist, may be observed in the new data collected during the
 RunII at the Tevatron and then at the LHC, or the limits on the
 mass of such a particles will be improved.

 \section*{\label{UHECR} Appendix 9. Neutral Tera-Mesons and Charged Tera-Leptons in  UHECR }

  In top-down model Ultra High Energy Cosmic Rays, (UHECR)
   are born by the decay of superheavy particles (e.g. topological
  defects) or their annihilations \cite{DFK}. These high energy sources will
  provide  an unique laboratory for tera leptons $E^-$,$E^+$ as
  well as  heavy quarks $U$ and $\bar{U}$ : these Sinister
 particles
will be produced at high energy (ZeV or above) along a tail
  of all possible exotic (as SUSY secondary ones \cite{Datta} )
 within UHECR spectra.
 Such a High energy Leptons pairs $E^-$,$E^+$  (at energy above
GZK cut-off) born in the far Universe edges has the very peculiar
behavior to escape along the space ignoring electromagnetic
interaction suffering negligible energy loss (contrary,for
instance, to UHECR proton or nucleons at GZK energies). This
ability to overcome both BBR and radio viscosity is based on the
huge ($m_E \simeq 500 GeV $) tera-lepton mass and its consequent
tiny Compton wave-length ($\lambda_E \propto\frac{1}{ m_E}$) and
its  consequent negligible electron pair production energy losses
as well as pion photo production. Also a negligible Tera-Lepton
photo-pion process is taking place for the same reasons described
below for Ultra High Energy  (UHE) Tera-Pions.
This allow to the stable (or nearly stable) UHE tera Leptons to
free travel inside the Earth and even cross the whole planet as
SUSY stable staus $\tilde{\tau}$
 \cite{Reno},  in full analogy to (the almost stable) UHE  lepton tau
 $\tau$ \cite{Fargion 2002} and  \cite{Fargion 2004} longer traces than muons
 ones. However out of very fine-tuned ad-hoc time-life these Tera-Leptons will be not able to
 produce Upward or Horizontal Tau-like Air-Showers, but just very
 long penetrating tracks with minor pairs production along the
 path.

   A different and more exciting role may come from the second
   Tera-Hadron UHE secondary of the
top-down UHECR source:
the birth of neutral Tera-Mesons $U\bar{u}$ and
   $u\bar{U}$. Contrary to early Universe, where there is enough
   time and hard dense matter to proceed in the very complex cooking of
   Tera-Hadrons relics summarized in present article, tera quarks
   born by UHECR source escape mostly as neutral $U\bar{u}$ and
   $u\bar{U}$.  These hybrid pions are  very exceptional candidates to UHECR  because
    two surprising abilities. The first is that  they interact hadronically  with
     matter.  This may imply a high altitude atmosphere interaction
     on Earth  in agreement with the observed UHECR fluorescence light curve.
       The second ability is a very smart interaction with radiation: indeed
       while the photo-pion production with matter takes place as
        for all the common nucleons (either proton or neutrons),
        the $threshold$ in the  "center of mass frame"  for tera-pions is much higher.
         The photo-pion production for neutrons onto the $2.75 K^o$ BBR begins
         at Lorentz factor $\gamma_{n \gamma} \simeq 4\cdot  10^{10}$ and energy
          $E_{n \gamma} \simeq 4\cdot  10^{19} eV$ while
         for a corresponding Lorentz factor the Hybrid Tera-Pion  threshold energy
          $E_ {U\bar{u} \gamma}\simeq 1.4\cdot 10^{23}\cdot S_6
         eV$, is   above three order of magnitude higher
         simply because of the huge Tera-pion mass.
         This imply that for Tera-Pion there is  a GZK cut
         off much above common  \cite{Greisen66},\cite{Zatsepin66}
         energies $E_{n \gamma} \simeq 4\cdot  10^{19} eV$
         making this rare UHE tera-pion an ideal candidate
         able to travel along the whole Universe without
         deflection of Galactic or Extra-galactic magnetic fields
         or much energy losses,  overcoming  the otherwise
           controversial absence
         of a  GZK  cut-off. These Tera-pions are among the best
         candidate  to solve recent   UHECR correlated events
         with far BL-Lac sources \cite{Gorbunov et al 2005}, \cite{HiRes}.



\begin{thebibliography}{99}

\bibitem[Agaronian and Vardanian 1985]{AV} Agaronian F.A., Vardanian V.V., 1985,
Preprint EFI-827-54-85-YEREVAN.


\bibitem[Belotsky et al 2000]{Sakhenhance}
Belotsky K.M., Khlopov M.Yu., Shibaev K.I., 2000, Gravitation and
Cosmology \textbf{6}, Suppl., 140.

\bibitem[Belotsky and Khlopov 2001]{BKS} Belotsky  K.M., Khlopov M.Yu., 2001,
Gravitation and Cosmology \textbf{7}, 189.

\bibitem[Belotsky and Khlopov 2002]{Belotsky} Belotsky  K.M., Khlopov M.Yu., 2002,
Gravitation and Cosmology \textbf{8}, Suppl., 112.

\bibitem[Belotsky Fargion Khlopov et al 2004]{4had}Belotsky K.M.,Fargion D.,
Khlopov M.Yu.,Konoplich R.V., Ryskin M.G., Shibaev K.I., 2004,
hep-ph/0411271

\bibitem[Berezinsky et al Astrophysics of Cosmic Rays 1990]{Ginzburg}
Berezinsky V.S.  et al. Astrophysics of Cosmic Rays, North Holland, 1990.

\bibitem[Burns and  Lovelace 1982]{BL} Burns M.L., Lovelace R.V.E., 1982, Astrophys.J.  \textbf{202}, 87.

\bibitem [Datta Fargion Mele 2005]{Datta}Datta A., Fargion D., Mele B., 2004,
hep-ph/0410176, in press.

\bibitem[Dolgov 2002]{Dolgov}
Dolgov A.D., 2002, Phys.Rept. {\bf 370}, 333; hep-ph/0202122.

\bibitem[Dubrovich Fargion  Khlopov 2004]{DFK} Dubrovich V.K., Fargion D., and Khlopov M.Yu., 2004, Astropart. Phys.  \textbf{22}, 183;  hep-ph/0312105.

\bibitem[Eidelman et al 2004]{PDG} Eidelman S. et al.,(Particle Data Group), 2004, Phys. Lett.
\textbf{B592}, 1.


\bibitem[Fargion et al 1999]{Fargion99} Fargion  D. et al, 1999, JETP Letters \textbf{69}, 434;
astro/ph-9903086

\bibitem[Fargion et al 2000]{Grossi} Fargion  D. et al, 2000, Astropart. Phys. \textbf{12}, 307;
astro-ph/9902327

\bibitem[Fargion 2002]{Fargion 2002}Fargion D., 2002, Astrophys.J. \textbf{570}, 909; astro-ph/0002453.

\bibitem[Fargion et al 2004]{Fargion 2004} Fargion D., De Sanctis Lucentini P.G., De Santis M., Grossi M., 2004, Astrophys.J. \textbf{613} 1285.

\bibitem[Glashow 2005]{Glashow} Glashow S.L., 2005,  hep-ph/0504287; Cohen A.G. and Glashow S.L., in preparation.
\bibitem[Gorbunov et al 2005]{Gorbunov et al 2005}  Gorbunov  D. S., Tinyakov  P. G., Tkachev I. I. ,
Troitsky S. V., 2005,  astro-ph/0505597; Gorbunov  D. S., Tinyakov
P.,2005, Astropart.Phys.\textbf{ 23},  175.

\bibitem[Greisen1966]{Greisen66} Greisen, K., 1966, Phys. Rev. Lett., \textbf{16}, 748

\bibitem[Hemmick et al 1990]{exph2} Hemmick T.K. et al.,1990, Phys. Rev. \textbf{D41}, 2074.

\bibitem[HiRes Collaboration 2005]{HiRes} HiRes Collaboration 2005; astro-ph/0507120


\bibitem[Klein et al 1981]{exp1} Klein J. et al., 1981, in \textit{Proceedings of the Symposium on
Accelerator Mass Spectrometry (Argonne National Laboratory,
Argonne, IL, 1981)}.


\bibitem[ Khlopov 1981]{fractons} Khlopov M.Yu., 1981, JETP Lett. {\bf 33}, 162.


\bibitem[Khlopov Cosmoparticle physics 1999]{book} Khlopov M.Yu., Cosmoparticle physics, World Scientific, 1999.



\bibitem[Khlopov and Linde 1984]{Linde} Khlopov M.Yu., Linde A.D.,1984,
 Phys. Lett. \textbf{138B}, 265; Balestra F. et al., 1984, Sov.J.Nucl.Phys.
 \textbf{39}, 646; Nuovo Cim. \textbf{A79}, 193;
Khlopov M.Yu. et al., 1994, Phys.Atom.Nucl. \textbf{57},  1393.

\bibitem[Khlopov and Shibaev 2002]{Shibaev} Khlopov M.Yu., Shibaev K.I., 2002,
Gravitation and Cosmology \textbf{8}, Suppl., 45.

\bibitem[Maltoni et al 2000]{Okun} Maltoni M. et al., 2000, Phys. Lett. {\bf B476},
107; Ilyin V.A. et al., 2001, Phys. Lett. {\bf B503}, 126;
Novikov V.A. et al., 2002, Phys. Lett. {\bf B529}, 111; JETP Lett.
{\bf 76}, 119.

\bibitem[Middleton et al 1979]{exph1} Middleton R. et al., 1979, Phys. Rev. Lett. \textbf{43}, 429.


\bibitem[Mueller et al 2004]{exp3} Mueller P. et al., 2004, Phys. Rev. Lett. \textbf{92}, 022501.

\bibitem[  Reno et al 2005] {Reno} Reno M. H., Sarcevic I., Su S., 2005, hep-ph/0503030.  See also Dutta S. I., Huang Y., Reno M.H., 2005, hep-ph/0504208.

\bibitem[Scherrer and M. Turner 1986]{Turner}
Scherrer R. and Turner M., 1986, Phys.Rev. {\bf D 33}, 1585.


\bibitem[Smith et al 1982]{exph3} Smith P.F. et al., 1982, Nucl. Phys. \textbf{B206}, 333.

\bibitem[Vandegriff et al 1996]{exp2} Vandegriff J. et al., 1996, Phys. Lett. \textbf{B365}, 418.

\bibitem[Zatsepin Kuz'min1966]{Zatsepin66} Zatsepin, G.T., \& Kuz'min, V.A., 1966, JETP Lett. \textbf{4}, 78.

\bibitem[Zeldovich and  Novikov  1983]{ZeldNov}
Zeldovich Ya.B. and Novikov I.D., Struktura i Evolyutsiya
Vselennoi (in Russian), Nauka, Moscow, 1975; Structure and
Evolution of the Universe, The University of Chicago Press, 1983.


\bibitem[Zeldovich and Khlopov 1979]{ZK} Zeldovich Ya.B., Khlopov M.Yu., 1978, Phys. Lett. {\bf B79}, 239.
\end{thebibliography}
\end{document}